%% file: ms.tex
\documentclass[journal]{IEEEtran}

\usepackage{setspace}
\usepackage{xcolor}
\usepackage{graphicx}
\usepackage{multirow}
\usepackage{booktabs}
\usepackage{subfig}
\usepackage{pifont}
\usepackage{algorithm}
\usepackage{algorithmic}
\usepackage{tabularx}

\let\oldding\ding
\renewcommand{\ding}[2][1]{\scalebox{#1}{\oldding{#2}}}

\newcommand\rev[1]{{#1}}
\newcommand\revA[1]{{#1}}
\newcommand\revB[1]{{#1}}
\newcommand\revC[1]{{#1}}

\usepackage{cite}

\begin{document}

\title{MoRS: An Approximate Fault \underline{Mo}delling Framework for \underline{R}educed-Voltage \underline{S}RAMs}

\author{
    \IEEEauthorblockN{\.{I}smail Emir Y\"{u}ksel\IEEEauthorrefmark{3}, Behzad Salami\IEEEauthorrefmark{2},  O\u{g}uz Ergin\IEEEauthorrefmark{3}, Osman S. \"{U}nsal\IEEEauthorrefmark{2}, Adri\'{a}n Cristal Kestelman\IEEEauthorrefmark{2}\IEEEauthorrefmark{4}}

   \IEEEauthorblockA{\IEEEauthorrefmark{3} TOBB University of Economics and Technology (TOBB ET\"{U})}
   \IEEEauthorblockA{\IEEEauthorrefmark{2} Barcelona Supercomputing Center (BSC)}
   \IEEEauthorblockA{\IEEEauthorrefmark{4} Universitat Politècnica de Catalunya (UPC)}
}

\maketitle

\input{sections/00-abstract}


\IEEEpeerreviewmaketitle

\input{sections/01-introduction}

\input{sections/02-background}

\input{sections/03-mors_framework}

\input{sections/04-ex_methodology}

\input{sections/05-ex_results}
\input{sections/06-related_work}
\input{sections/07-conclusion}
\input{sections/08-ack}

\bibliographystyle{IEEEtran}
\bibliography{references}
\input{sections/08-authors_biography}


%\vfill


\end{document}

%% file: sections/00-abstract.tex
\begin{abstract}
    On-chip memory (usually based on Static RAMs-SRAMs) are crucial components for various computing devices including heterogeneous devices, \textit{e.g,} GPUs, FPGAs, ASICs to achieve high performance. 
    Modern workloads such as Deep Neural Networks (DNNs) running on these heterogeneous fabrics are highly dependent on the on-chip memory architecture for efficient acceleration.
    Hence, improving the energy-efficiency of such memories directly leads to an efficient system. 
    One of the common methods to save energy is undervolting \textit{i.e.,} supply voltage underscaling below the nominal level. 
    Such systems can be safely undervolted without incurring faults down to a certain voltage limit.
    This safe range is also called voltage guardband.
    However, reducing voltage below the guardband level without decreasing frequency causes timing-based faults.

    In this paper, we propose MoRS, a framework that generates the first approximate undervolting fault model using real faults extracted from experimental undervolting studies on SRAMs to build the model. We inject the faults generated by MoRS into the on-chip memory of the DNN accelerator to evaluate the resilience of the system under the test. MoRS has the advantage of simplicity without any need for high-time overhead experiments while being accurate enough in comparison to a fully randomly-generated fault injection approach. We evaluate our experiment in popular DNN workloads by mapping weights to SRAMs and measure the accuracy difference between the output of the MoRS and the real data. \revA{Our results show that the maximum difference between real fault data and the output fault model of MoRS is 6.21\%, whereas the maximum difference between real data and random fault injection model is 23.2\%. In terms of average proximity to the real data, the output of MoRS outperforms the random fault injection approach by 3.21x. }

\end{abstract}

%% file: sections/01-introduction.tex
\section{Introduction}
\IEEEPARstart{S}{RAMs} are traditionally the building block of different components in different computing systems such as branch predictor (in CPUs), register files (in GPUs), on-chip buffer memory (in hardware accelerators like FPGAs and ASICs), thanks to the low-latency access time of such memories. However, the power consumption of SRAMs significantly contributes to the total system power. For instance, prior works on GPUs\cite{gpuRegFile1} show that register file in GPUs consumes 15-20\% of the total power. Another work on modern out-of-order cores\cite{oooEnergy} estimates that the SRAM-dominated front-end of the microprocessor consumes up to 33\%  of the total power of the CPU. Other work on DNN accelerators on FPGAs \cite{salami2018comprehensive} shows that on-chip SRAMs consume 27\% of the total power. 

Since the total power consumption of any underlying hardware is directly related to its supply voltage, voltage underscaling is an effective solution to save power~\cite{papadimitriou2020exceeding, gizopoulos2019modern}. This technique is widely used in various devices such as CPUs\cite{bacha2014using,papadimitriou2019adaptive,parasyris2018aframework,papadimitriou2017harnessing,bacha2013dynamic}, GPUs\cite{zou2018voltage} FPGAs\cite{salami2018comprehensive,salami2020experimental, salami2019evaluating, salami2018fault, salami2018aggressive}, and ASICs\cite{reagen2016minerva,chandramoorthy2019resilient,zhang2018thundervolt} as well as  DRAMs\cite{chang2017understanding,david2011memory,deng2011memscale}, HBMs\cite{larimi2020understanding}, SRAMs\cite{yang2017sram,Yang2017ApproximateSF}, and Flash Disks\cite{cai2013error,caierror2017,cai2012error,cai2015read}. However, while performing voltage underscaling, reliability issues can arise due to the increased circuit delay at reduced voltage levels. In most commercial devices, there is a timing-fault free guardband between nominal voltage and minimum safe voltage, \textit{i.e.,} $V_{min}$. Below this voltage guardband, faults occur as a consequence of circuit delays. 
%While supply voltage decreases, there is a voltage crash level that hardware stops operating. 

Although further aggressive voltage underscaling below $V_{min}$ can achieve more reduction in power consumption, it compromises system reliability through undervolting faults. This particular region between $V_{min}$ and the lowest operational voltage level, \textit{i.e.,} $V_{crash}$, still operational but with faults is called the critical area\cite{salami2018comprehensive}. Several prior fault mitigation techniques were proposed\cite{salami2018comprehensive,reagen2016minerva,chandramoorthy2019resilient,torreshuitzil2017fault,deng2015retraining}. However, these techniques need either high-effort engineering\cite{salami2018comprehensive,deng2015retraining} or totally random fault injection campaigns\cite{reagen2016minerva,chandramoorthy2019resilient}. These approaches are either impractical or not accurate. Our solution offers the advantages of both random fault injection and empirical experiments.

In this study, we propose MoRS, a framework that generates the first approximate voltage underscaled model for SRAMs. MoRS consists of three steps: 1) Experiment, 2) Behavior Extraction, and 3) Model Generation. In the Experiment step, MoRS uses publicly available undervolted fault map data\cite{salami2018bram} of SRAMs based on our prior work\cite{salami2018comprehensive}. In the Behavior Extraction step, we extract the characteristic fault behavior features of undervolted SRAM blocks. We establish and confirm that undervolting based faults do not occur randomly. These faults are correlated with each other in space. We examine the faults of row-based and column-based approaches and see the distance between consecutively faulty bitcells, the number of each bit-fault in both rows and columns, and the total number of faulty rows and columns per SRAM block are not uniformly distributed. These fine-grained features show characteristic behaviors. In this step, we extract characteristic features and categorize them into two profiles: coarse-grained and fine-grained profiling. The reason behind these categorizing is that random fault injection studies use only coarse-grained features, the number of bit faults, and the number of faulty SRAM blocks. The output of MoRS that we term Mixed Model uses both fine-grained and coarse-grained features and applies probabilistic custom modeling algorithm to achieve an approximate model as the last step.

In recent years, the power consumption of SRAMs on DNNs is increasing drastically\cite{reagen2016minerva,salami2018comprehensive,chandramoorthy2019resilient}. We evaluate MoRS for DNN accelerators where on-chip SRAMs play an important role. MoRS generates fault models that are not limited to a certain domain and can be potentially used in many domains such as branch prediction units, register files, caches, and any other memories based on SRAM, unlike the other prior studies. To evaluate MoRS, we generate a baseline model by applying a uniform random distribution function to coarse-grained features. 
%This baseline model is mainly used in prior studies\cite{chatzidimitrou2018analysis,chandramoorthy2019resilient,reagen2016minerva,salami2018ontheresilience}, as a fault injection model, we also called Random Model. 
This baseline, which we term as Random Model, is a standard fault injection scheme and used in prior studies\cite{chatzidimitrou2018analysis,chandramoorthy2019resilient,reagen2016minerva,salami2018ontheresilience}.
Besides, we process empirical data to see the difference in accuracy between each artificial model (Random Model and Mixed Model) and real data. 

In our evaluation methodology, we map weights to SRAM blocks.
When we map weights to SRAM blocks, there can be different mapping options such as MSB mapping, LSB mapping, the first half of bits MSB the other half of bits LSB mapping, and the first half of bits LSB the other half of bits MSB mapping. Another method to save energy on DNNs is quantization. Our quantization method is reducing the precision of weights from 32-bit to 16-bit, 8-bit, 4-bit, and 1-bit, respectively. While performing undervolting, some unwanted bits can be flipped, and then the corresponding value becomes infinity or NaN. To avoid these values masking techniques are used\cite{koppula2019eden}. We mask infinity or NaN value to 1 or 0. We examine the behavior of different weight mappings, quantizations, and masking options. We evaluate our experiments on trained LeNeT-5\cite{lecun1998lenet} and cuda-convnet\cite{krizhevsky2014cuda} DNNs in the classification stage. We examine the classification accuracy for each voltage level. Our experiments show that generated artificial model has similar behavior with the real data on different DNN benchmarks with an accuracy of 96.4\%. We also see that Random Model is not close enough with real data with the difference in accuracy up to 23\%. We find that on average our Mixed Model has 3.6\% difference with real data and on average 3x up to 7x closer than the baseline random model. 

Our contributions are as follows:
\begin{itemize}
\item We propose a framework, MoRS that generates an artificial model that can realistically emulate real undervolting fault data with a difference in the accuracy of 3.6\% on average. To the best of our knowledge, this study provides the first reasonably accurate model for SRAM blocks under low-voltage conditions.  
\item We evaluate our models and real data on state-of-the-art DNNs to see how different weight mappings, quantization, and value maskings affect the accuracy of inference. We find \revA{the similar observation with prior works\cite{sabbagh2019evaluating,guanpeng2017understanding}} that if we continue to reduce the precision of weights, DNNs become more resilient to undervolting. At the lowest reduced voltage level 8-bit LeNeT accuracy is 14\% while 4-bit LeNeT accuracy is 60\%. Even in this unlikely situation, our Mixed Model shows similar behavior to real data.
\end{itemize}
The remainder of this paper is structured as follows. In Section II, we introduce the most important concepts used in this paper. In Section III, we propose our approximate fault modeling framework, MoRS. Section IV describes the methodology that we perform our experiments. Section V detail the experimental results. Related works are introduced in Section VI. Finally, Section VII concludes this paper.

%% file: sections/02-background.tex
\section{Background}
\subsection{Undervolting SRAM based on-chip Memories}
CMOS is the dominant circuit technology for current computing devices. 
The power consumption of CMOS-based SRAMs is the sum of two parts: the first is dynamic power, also called active power, and the other one is leakage power, also called static power. Earlier studies\cite{updahyay2014low,azam2010variability,chen2010yield} show that the power consumption of SRAMs is dominated by dynamic power.

Dynamic power \textit{i.e.,} $P_{dyn}$, is dissipated when there is switching activity at some nodes in transistors. 
Leakage power \textit{i.e.,} $P_{leak}$, is dissipated by the leakage currents flowing even when the device is not active.
Mathematically, these power equations are given by,
\begin{equation}
    P_{dyn} \propto  f \times V_{dd}^{2} 
    \label{eq:dyn_p}
\end{equation}
\begin{equation}
    P_{leak} = k_{design} \times n  \times I_{leak} \times V_{dd}
    \label{eq:leak_p}
\end{equation}

Here $V_{dd}$ shows the supply voltage, $f$ shows the operating frequency in Equation \ref{eq:dyn_p}. Further, $n$ denotes the number of transistors, $k_{design}$ is the design-dependent parameter, and $I_{leak}$ shows the leakage current a technology-dependent parameter in Equation \ref{eq:leak_p}. 
From Equation \ref{eq:dyn_p}, dynamic power consumption can be reduced by adjusting the supply voltage and operating frequency. Likewise, from Equation \ref{eq:leak_p}, leakage power consumption can be reduced by underscaling the supply voltage and reducing the total number of transistors. Total power consumption can be reduced by underscaling the supply voltage. 

Voltage underscaling is a widely used technique for energy-efficient applications. From Equation \ref{eq:dyn_p}, dynamic power consumption is reduced quadratically by underscaling the supply voltage. This technique can achieve nominal voltage level \textit{i.e.,} $V_{nom}$ performance due to the operating frequency not being changed. However, further aggressive undervolting below the minimum safe voltage \textit{i.e.,} $V_{min}$ may cause reliability issues as the result of timing faults. Between $V_{nom}$ and $V_{min}$ is called voltage guardband. This guardband is to provide an assurance of correct functionality even in the worst environmental case. Prior works achieve by applying undervolting that power consumption reduces; 39\% on FPGA on-chip memories~\cite{salami2018comprehensive}, 20\% in GPUs~\cite{gpuSafeLmits}, and 16\% in DRAMs~\cite{chang2017understanding} without any timing-related errors. Also, recent studies show that undervolting internal components of FPGAs~\cite{salami2020experimental} lead to around 3x power-efficiency and underscaling supply voltage of HBMs~\cite{larimi2020understanding} achieve a total of 2.3x power savings. In Minerva\cite{reagen2016minerva}, lowering SRAM voltages achieves a total of 2.7x power savings.

\subsection{Fault Injection}
The fault injection mechanism is a way to examine the behavior of systems under different circumstances. Prior works on fault injection study for reduced-voltage SRAMs are applied on branch prediction units~\cite{chatzidimitriou2019assessing,chatzidimitrou2018analysis}, caches~\cite{riviere2015high,kailorakis2015differential}, and FPGA on-chip memories~\cite{salami2018ontheresilience}. \revA{There are several approaches for the fault injection with having a trade-off on the engineering effort and accuracy.} i) The first one is applying random faults to random locations without any information used from empirical experiments. ii) Another one is directly using the empirical data as a fault map. iii) The last one is the approximate modeling that is based on real data from empirical experiments and close enough to empirical data.  Table \ref{tab:method_comparison} gives a summary of comparison of these three fault injection techniques in terms of effort and how close they are to real data. The accuracy percentage is from the results of this study.

% \usepackage{booktabs}
\begin{table}[h]
\centering
\caption{Comparison of fault injection techniques in terms of engineering effort and accuracy of representing real data }
\label{tab:method_comparison}
\resizebox{\linewidth}{!}{%
\begin{tabular}{@{}ccc@{}}
\toprule
\multicolumn{1}{l}{\textbf{Fault Injection Method}} & \multicolumn{1}{l}{\textbf{Engineering Effort}} & \multicolumn{1}{l}{\textbf{Accuracy (Min)}} \\ \midrule
Random Fault Injection & Low       & 77\%  \\
\rev{\textbf{MoRS}}  & \rev{Low}    & \rev{94\%}  \\
Empirical Data         & High & 100\% \\ \bottomrule
\end{tabular}%
}
\end{table}
\subsection{Deep Neural Networks}
Deep Neural Networks (DNNs) are widely used as an effective solution for object recognition, classification and segmentation, and many other areas. MNIST\cite{mnist} and CIFAR-10 \cite{cifar10} datasets are widely used by the ML community to showcase the latest technology advancements in DNNs. DNNs perform in two phases: training and prediction (inference). The first process is training that weights of DNNs are fitted to data. The training phase is mainly performed in high-performance computing platforms like CPUs or GPUs. The second phase is the inference phase, DNN models, trained in the first phase, are used to extract unknown output from input data. Usually, the training phase performs once whereas the prediction phase performs repeatedly. The inference phase consists of mainly three elements: input, weights, and output. 
\begin{figure}[h]
  \centering
  \includegraphics[width=0.9\linewidth]{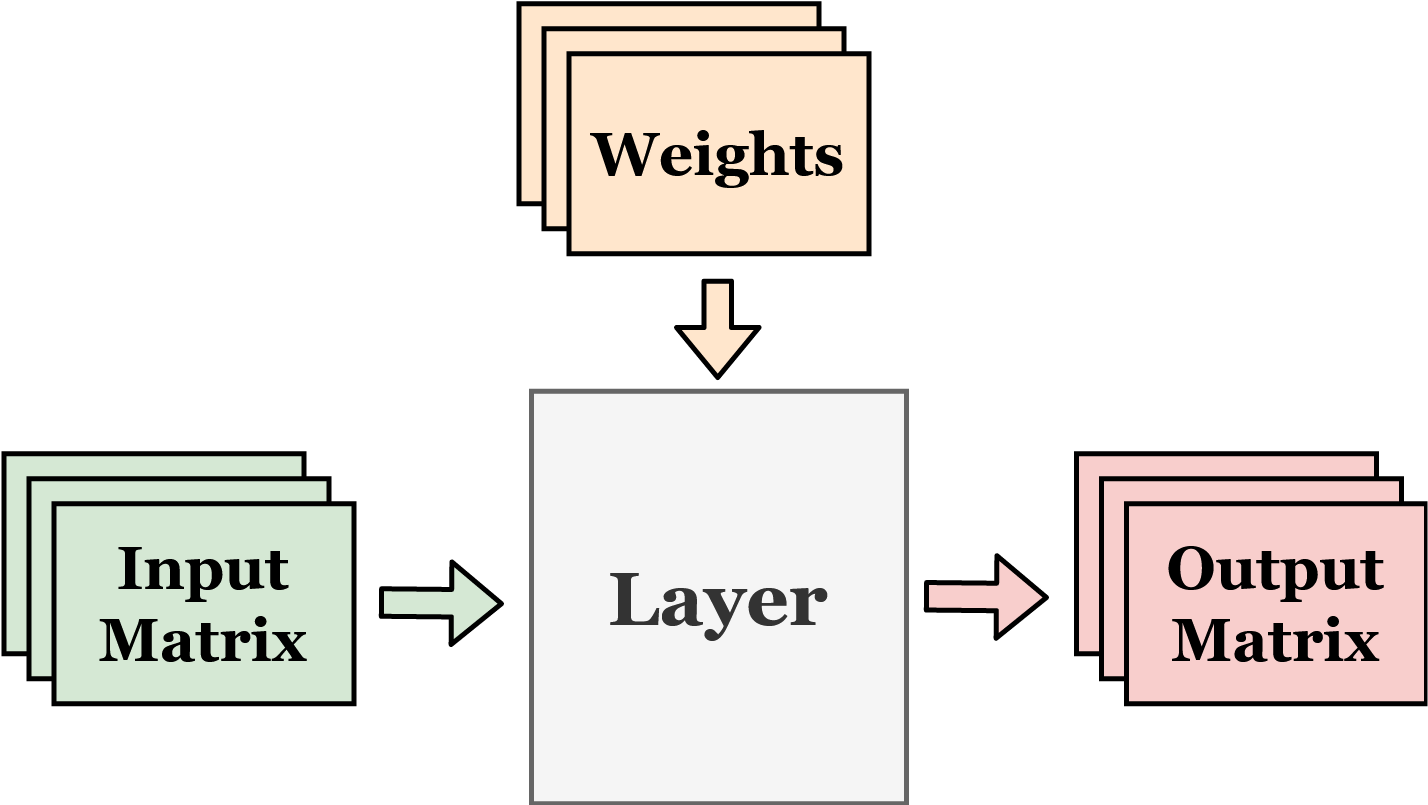}
  \caption{A high level abstraction of one layer of DNNs.}
  \label{fig:dnn_ex}
\end{figure}

Figure \ref{fig:dnn_ex} shows a high-level abstraction of one inference layer. Each inference layer of the network consists of weights, an input matrix, and an output matrix. Weights of these layers are stored in on-chip memories. Due to their large size and computational intensity, DNNs consume power; while they are fault-tolerant to some extend. Some studies improve the power-efficiency and reduce the total power consumption of DNNs by applying voltage underscaling \cite{salami2018comprehensive,salami2020experimental}, architectural techniques\cite{zhou2017incremental,han2016eie,zhubitrram2019,molchanov2017pruning,yazdanipruning2018,han2015learning,shenflexible2017,deng2020permdnn}, and hardware-level techniques\cite{shen2018maximizing,zhangoptimizing2015,riera2018computation}.

Quantization is an architectural optimization technique that reduces the precision of data types. This technique can improve performance and energy-efficiency. Recent studies\cite{courbariaux2016binarynet,reagen2016minerva,jacob2017quantization,wu2015quantized,ueyoshi2018quest} show that quantization to some extent does not significantly affect the accuracy of DNN models. In our study, we reduce weights precisions to 16-bit half-precision floating point, 8-bit($Q_{4.4}$), 4-bit($Q_{2.2}$) in fixed-point format, and 1-bit binary values.

%% file: sections/03-mors_framework.tex
\section{MoRS Framework} 
We propose MoRS, a framework to generate approximate artificial reduced-voltage SRAM fault models. This mechanism is the first framework based on real fault maps. MoRS stands between fully hardware and fully software fault injection techniques. This framework generates an approximate model that is close enough to real data compared to the fully software fault injection mechanism. Also, MoRS does not require high-effort engineering in comparison to a fully hardware approach that uses real data from empirical experiments.

\begin{figure}[h]
  \centering
  \includegraphics[width=\linewidth]{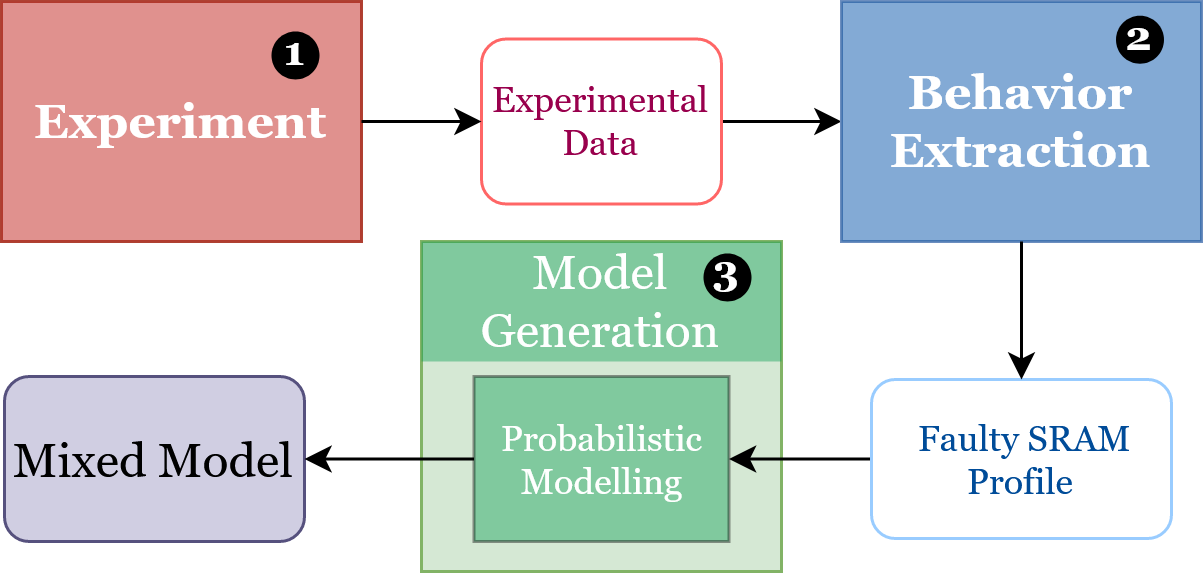}
  \caption{Overview of the MoRS}
  \label{fig:mechanism}
\end{figure}

As shown in Figure \ref{fig:mechanism} MoRS consists of three steps: \ding[1.5]{182} Experiment, \ding[1.5]{183} Behavior Extraction, and \ding[1.5]{184} Model Generation. We first explain the first step existing experiment that provides real fault maps from real SRAM blocks in Section 3.1. In Section 3.2 we extract the behavior of fault maps as fine-grained and coarse-grained profiles by using the output of the first step. In fine-grained profiling extract the row and column behaviors of SRAM blocks in terms of physical distance, the number of each bit-faults in rows and columns, and the number of faulty rows and columns per block. Coarse-grained profiling is a shallow approach generally used in prior fault injection studies. In coarse-grained profiling, we extract only the total number of bit-faults and faulty SRAM blocks. In Section 3.3 we generate an artificial model with outputs of the second steps. 

At the endpoint, we provide Mixed Model which is an approximate model to real data. It is generated by applying probabilistic modeling on both fine-grained and coarse-grained fault profiles. Figure \ref{fig:mechanism} provides an overview of the three steps of the MoRS.  
\subsection{Experiment}
\begin{figure}[!h]
  \centering
  \includegraphics[width=\linewidth]{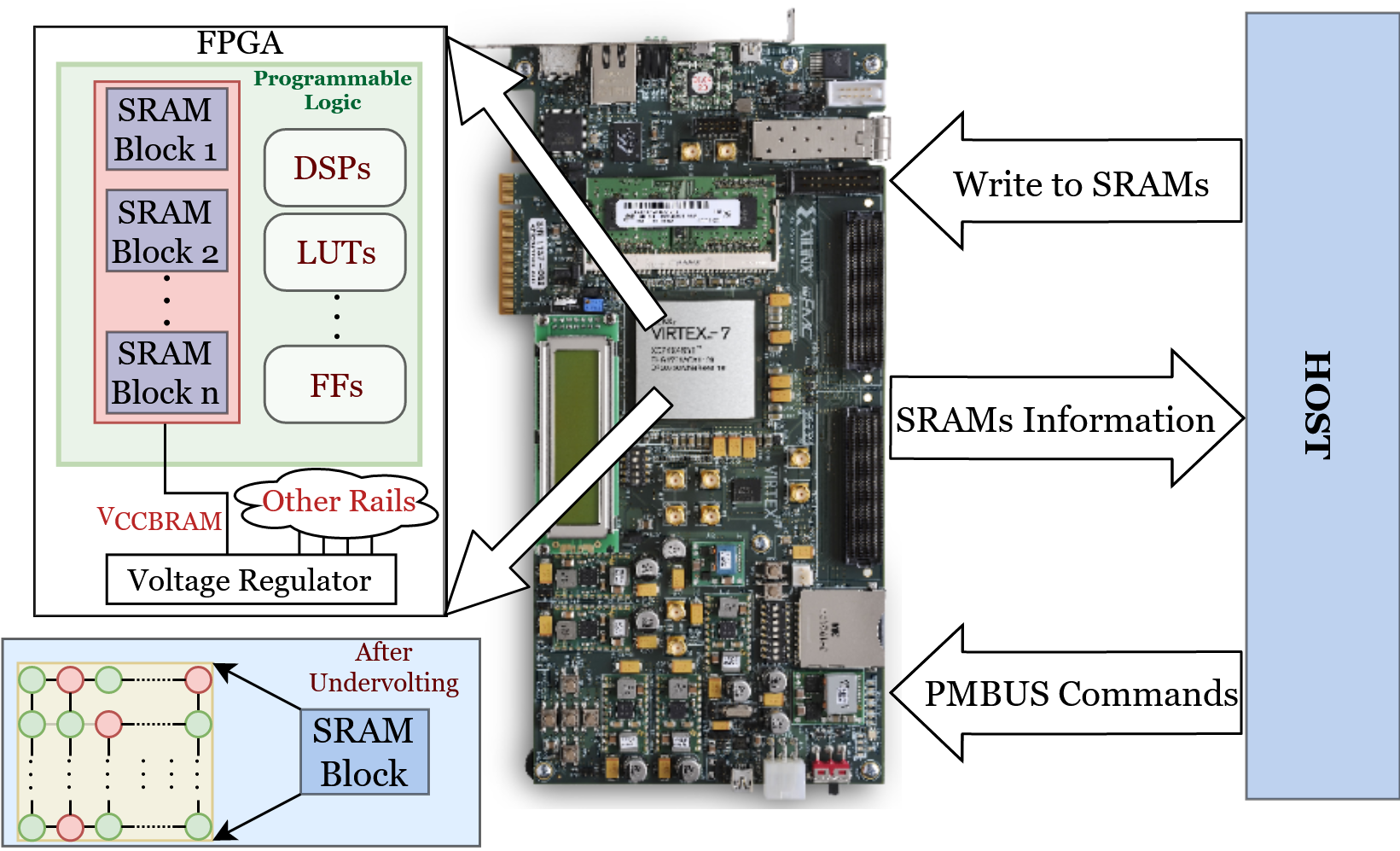}
  \caption{The methodology of empirical experiment}
  \label{fig:empirical}
\end{figure}

Voltage underscaling, \textit{i.e.,} undervolting is a widely used technique to save energy. We study undervolting for modern SRAM blocks. Each SRAM block is a matrix of bitcells formed of rows and columns. Also, each SRAM block is 16 Kbits with 1024 rows and 16 columns. 
\begin{table}[]

\begin{tabular}{|l|c|c|c|}
\hline
\multicolumn{1}{|c|}{\textbf{Board Name}} & \textbf{VC707} & \textbf{ZC702} & \textbf{KC705} \\ \hline\hline
\textbf{Technology Node}       & 28nm  & 28nm  & 28nm  \\ \hline
\textbf{Nominal Voltage}       & 1V    & 1V    & 1V    \\ \hline
\textbf{Operating Temperature} & $50^\circ C$    & $50^\circ C$    & $50^\circ C$    \\ \hline
\textbf{Minimum Voltage Level}                  & 0.54V & 0.53V & 0.54V \\ \hline
\textbf{Number of SRAM Block}      & 2060  & 280   & 890   \\ \hline
\end{tabular}

\caption{\revC{The summary of deployed FPGA boards}}
\label{tab:fpga_detail}
\end{table}
\revB{We perform this empirical study on SRAM blocks available in the off-the-shelf Xilinx FPGAs, i.e., VC707, ZC702, and two identical samples of KC705 (referred as KC705-A and KC705-B).} \revC{Table \ref{tab:fpga_detail} shows the detail of deployed FPGAs.} This experiment consists of two parts. The first part is the monitor SRAM blocks to see undervolting faults. 
The second part is adjusting the supply voltage of SRAM blocks by using a power management unit of FPGAs. FPGAs have a power management bus (PMBUS) a voltage regulator that monitors and adjusts voltage rails on FPGAs. This bus uses PMBUS standard and has an I2C protocol to execute commands. The supply voltage of SRAM blocks of FPGAs is named $V_{CCBRAM}$. By reducing $V_{CCBRAM}$ through PMBUS, only the power consumption of SRAM blocks reduces. Since this situation does not affect any other logic parts (DSPs, LUTs, etc..) the effect of undervolting on SRAM blocks is seen clearly. The setup of this methodology is shown in Figure \ref{fig:empirical}.
The method of this experiment follows an algorithm that as a first step writes data to SRAMs after in the second step analyzes the faults in terms of rate and location, in the third step reduces the supply voltage by 10$mV$, finally repeat these steps until FPGA crashes.

When undervolting is applied below the voltage guardband \textit{i.e.,} $V_{min}$, the fault rate exponentially increases. Voltage can be reduced until the voltage that FPGA stops operating \textit{i.e.,} $V_{crash}$. Between $V_{min}$ and $V_{crash}$ faults occur due to timing failures while the power consumption significantly reduces. SRAM-based real fault maps significantly vary even for different FPGAs of the same platform as the result of the process variation. Besides, the study shows that the pattern of faults is mostly permanent. Also, we observe that the fault rate and location for different run are mostly the same. Most importantly, undervolting faults are not uniformly distributed over different SRAMs. 

We use the publicly available data\cite{salami2018bram} of the prior work \cite{salami2018comprehensive} as the output of the first stage (Experimental Data) to generate artificial fault maps. We use 2000 SRAM blocks for extracting behavior and 950 SRAM blocks for testing our framework, 2060 SRAM blocks from VC707 and 890 SRAM blocks from KC705. We make the methodology of MoRS reliable by using different data set to create and also to test our artificial model.

At $V_{crash}$ the fault rate, up to 0.06\% and 0.005\% per 1 Mbits for VC707 and KC705-B, respectively. It should be noted that faults appear in between $V_{min}$ = 0.6V, $V_{crash}$ = 0.54V for VC707 and $V_{min}$ = 0.59V, $V_{crash}$ = 0.53V for KC705-B. As the prior work\cite{salami2018comprehensive} mentioned VC707 has the most bit-faults among the other three boards. At the lowest voltage level, VC707 has 10.2\% faulty SRAMs with 23706 bit-faults. \rev{We refer \cite{salami2018comprehensive} to more detailed information about this prior work.} 
\begin{table*}[ht]
\caption{Features and which model they are used in}
\label{tab:feat}
\resizebox{\linewidth}{!}{%
\begin{tabular}{|c|l|l|c|c|}
\hline
\multirow{2}{*}{\textbf{Profiling Type}} & \multicolumn{2}{c|}{\textbf{Features}}                   & \multicolumn{2}{c|}{\textbf{Models}}                    \\ \cline{2-5} 
 &
  \textbf{Name} &
  \textbf{Percentage} &
  \multicolumn{1}{l|}{\textbf{Mixed-Model}} &
  \multicolumn{1}{l|}{\textbf{Random-Model}} \\ \hline
\multirow{2}{*}{\textbf{Coarse-grained}} &
  Total number of bit-faults &
  $P_{F}$ &
  \ding{51} &
  \ding{51} \\ \cline{2-5} 
                                         & Total number of faulty SRAM blocks        & $P_{S}$            & \ding{51} & \ding{51} \\ \hline
\multirow{6}{*}{\textbf{Fine-grained}} &
  Total number of faulty rows per SRAM block &
  $P_{SR_{0..1024}}$ &
  \ding{51} &
  \ding{55}  \\ \cline{2-5} 
                                         & Total number of faulty columns per SRAM block      & $P_{SC_{0..16}}$           & \ding{51} & \ding{55}  \\ \cline{2-5} 
                                         & Number of each bit-fault in rows    & $P_{FR_{0..15}}$   & \ding{51} & \ding{55}  \\ \cline{2-5} 
                                         & Number of each bit-fault in columns & $P_{FC_{0..1023}}$ & \ding{51} & \ding{55}  \\ \cline{2-5} 
 &
  Distance between consecutively faulty bitcells in rows &
  $P_{FDR_{1..15}}$ &
  \ding{51} &
  \ding{55}  \\ \cline{2-5} 
 &
  Distance between consecutively faulty bitcells in columns &
  $P_{FDC_{1..1023}}$ &
  \ding{51} &
  \ding{55} \\ \hline
\end{tabular}
}
\end{table*}
\subsection{Behavior Extraction}
As we mentioned in Section 3.1 voltage underscaling faults have patterns and these fault patterns \textit{i.e.,} fault maps, are mostly the same. \revB{There are features that affect the probability of bit-faults. In this step, we profile the behavior of undervolting-related bit-faults to extract such important features.} We perform the profiling in two steps: coarse-grained profiling and fine-grained profiling. A summary of all features, profiling types, and which model they are used in can be found in Table \ref{tab:feat}.

Coarse-grained profiling consists of two features. The first one is the percentage of bit-faults in all SRAMs bitcell \textit{i.e.,} $P_{F}$. The second one is the percentage of faulty SRAMs in all SRAM blocks \textit{i.e.,} $P_{S}$. For VC707, at $V_{crash}$, total bit-faults are 0.07\% and the percentage of faulty SRAMs is 10.2\%. 

Fine-grained profiling comprises two parts: row-based, column-based. Both row-based and column-based have three features. The first one is the percentage of faulty rows \textit{i.e.,} $P_{SR_{0..1024}}$ and faulty columns $P_{SC_{0..16}}$ in faulty SRAM blocks. The second one is the percentage of each bit-faults in rows \textit{i.e.,} $P_{FR_{0..16}}$, or columns \textit{i.e.,} $P_{FC_{0..1024}}$. The last one is the percentage of physical distance between consecutively faulty bitcells \textit{i.e.,} bitcell-distances, in the same row \textit{i.e.,} $P_{FDR_{1..15}}$, or column \textit{i.e.,} $P_{FDC_{1..1023}}$. We discover that these two features for row and column are not randomly or uniformly distributed. 

At $V_{crash}$,
\begin{itemize}
    \item The percentage of each bit-faults
     \begin{itemize}
         \item In terms of row concentrates at 2-bit faults ($P_{FR_{2}}$) and no faults ($P_{FR_{0}}$).
         \item In terms of column it concentrates no faults to 10-bit faults ($P_{FC_{0..10}}$).
     \end{itemize}
     \item The percentage of bitcell-distances between consecutively faulty bitcells
     \begin{itemize}
         \item In terms of row concentrates at 8-bit distance ($P_{FDR_{8}}$).
         \item For rows that there is no such bitcell-distance is more than 8-bitcells ($P_{FDR_{9..15}}$).
        
         \item For the columns, it is concentrated in even numbers in decreasing order$P_{FDC_{0..2..1024}}$, 2-bitcells distance ($P_{FDC_{2}}$) has the highest percentage while 1022-bitcells distance ($P_{FDC_{1022}}$) has the lowest in even numbers.
         \item Also, for columns, the bitcell-distances in odd numbers ($P_{FDC_{1..3..1023}}$) are stuck between 0 and 4.
     \end{itemize}
\end{itemize}

\begin{figure}[htp]
  \centering
  \includegraphics[width=0.85\linewidth]{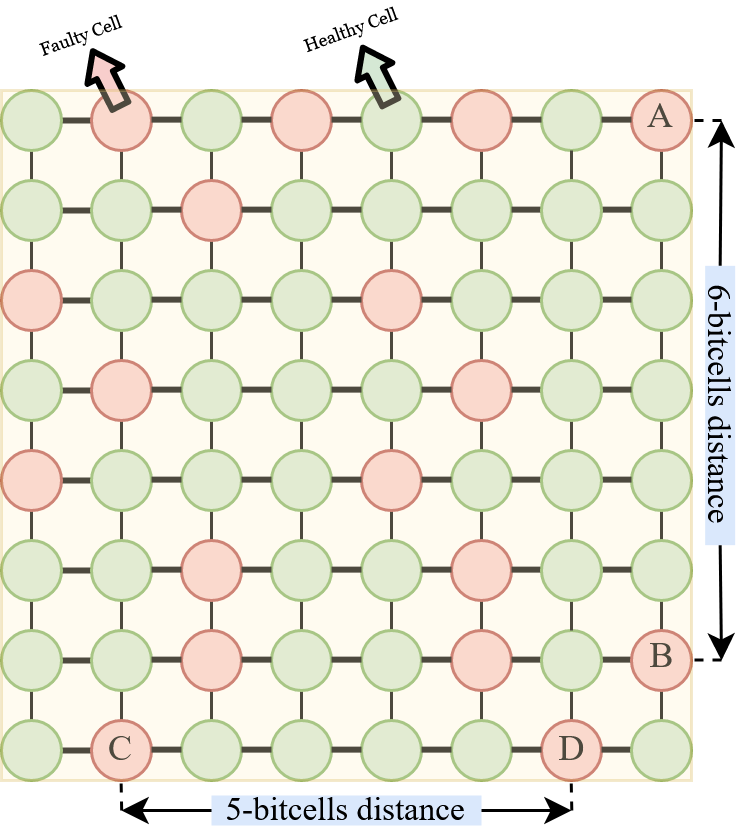}
  \caption{Physical features on a faulty SRAM block as an example}
  \label{fig:features_fig}
\end{figure}

As an illustration, we show the fault behavior of an $8*8$ SRAM block in Figure \ref{fig:features_fig}. In this example, the column-based distance between cell A and cell B is 6-bitcells. For cell C and cell D, the row-based distance is 5-bitcells. When we examine this example in Figure \ref{fig:features_fig} in terms of bit-fault for both row-based and column-based, a column that contains cell C has 2-bit faults. Also, cell A's row has 4-bit faults. One of the coarse-grained profile features is the percentage of total bit-faults ($P_{F}$). It can be calculated by dividing the total number of faulty cells by all cells. Hence, for Figure \ref{fig:features_fig}'s SRAM block $P_{F}$ is 28.125\%. 

As we extract the behavior of this example in Figure \ref{fig:features_fig}, we performed this process for experimental data. When the profiling step is done, we start generating models by using these coarse-grained and fine-grained profiles with probabilistic modeling and uniform random distribution.   
\begin{algorithm}
\caption{Generate Mixed Model}
\label{alg:mmpc}
\begin{algorithmic} 
\REQUIRE $n \leftarrow \#ofSRAMblocks$
\STATE $bitfaults \leftarrow n \times 1024 \times 16 \times P_{F}$
\STATE $faultyS \leftarrow n \times P_{S}$
\WHILE{$faultyS > 0 \ and \ bitfaults > 0$}
\STATE $block \leftarrow random$
\STATE $faultyBlock \leftarrow SRAMblocks[block]$
\STATE $faultyrows \leftarrow select from (F_{SR_{0..16}})$
\WHILE{$faultyrows > 0$}
\STATE $row \leftarrow random$
\STATE $column \leftarrow random$
\STATE $bitfaultsinrow \leftarrow select from (F_{FR_{0..16}})$
\WHILE{$bitfaultsinrow > 0$}
\STATE $distance \leftarrow select from (F_{FDR_{1..15}})$
\STATE $column \leftarrow column+distance$
\STATE $faultyBlock[row][column] \leftarrow fault$
\STATE $bitfaultsinrow \leftarrow bitfaultsinrow-1$
\ENDWHILE
\STATE $faultyrows \leftarrow faultyrows-1$
\ENDWHILE
\STATE $ArtificialCF \leftarrow ColumnFeatures(faultyBlock)$
\STATE $RealCF \leftarrow ColumnFeatures(RealData)$
\IF{$Similarity(ArtificialCF,RealCF) > 80\%$}
\STATE $SRAMblocks[block] \leftarrow faultyBlock$
\STATE $faultyS \leftarrow faultyS-1$
\STATE $bitfaults \leftarrow bitfaults-\#offaults(faultyBlock)$
\ENDIF
\ENDWHILE
\end{algorithmic}
\end{algorithm}
\subsection{Model Generation}
Probabilistic modeling \cite{koppula2019eden} and uniform random distribution \cite{koppula2019eden,chandramoorthy2019resilient,reagen2018ares,chatzidimitriou2019assessing} are widely used in many modeling studies to generate fault maps and to inject faults. For an approximate model \textit{i.e.,} Mixed Model, we use both fine-grained and coarse-grained features with custom probabilistic modeling function. In addition to these, we need the number of SRAM blocks that will generate.

To generate Mixed Model we follow the method shown in Algorithm \ref{alg:mmpc}. The input of Algorithm \ref{alg:mmpc} is the number of SRAM blocks and the output is faulty SRAM data, also called fault map. First, in Algorithm \ref{alg:mmpc}, faulty SRAM blocks are determined. After that, by using $P_{F}$ value, the number of faulty cells are calculated. First, we randomly select faulty SRAMs in all SRAM blocks. Then, we inject faults corresponding cells in faulty SRAMs. This injection algorithm uses fine-grained features. This process is performed in two stages: row-based fault injection and column-based control mechanism. The row-based fault injection uses row-based features. We determine how many rows will be faulty by using the probability of the number of faulty rows per SRAM block \textit{i.e.,} $F_{SR_{0..1024}}$ derivating from $P_{SR_{0..1024}}$. Then according to the probability of the number of each bit-faults in rows \textit{i.e.,} $F_{FR_{0..16}}$ derivating from  $P_{FR_{0..16}}$ we inject bit-faults to the corresponding row. To inject more than one fault in a row we use the probability of physical distance between consecutively faulty bitcells in rows $F_{FDR_{1..15}}$ derivating from $P_{FDR_{1..15}}$. In the second step of this algorithm, first, we extract column-based features of each artificial fault map. 

\begin{figure}[ht]
    \centering
    \includegraphics[width=\linewidth]{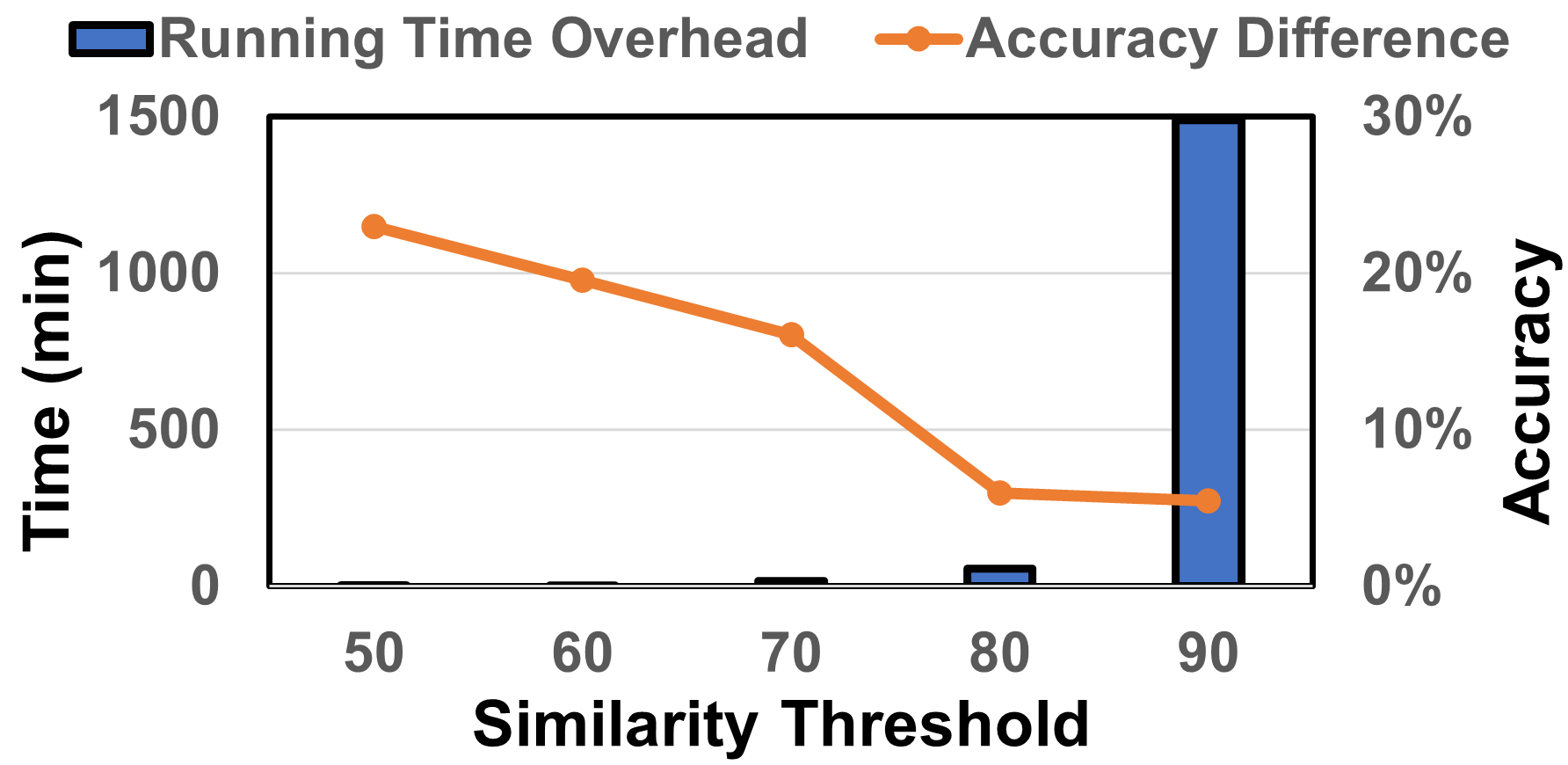}
    \caption{\rev{The running time overhead and accuracy difference between real data and the generated artificial model for different similarity threshold levels}}
    \label{fig:simi_threshold}
\end{figure}

\rev{After extraction, we compare these features with fine-grained column-based features extracted from the second step. If an artificial faulty SRAM has a lower than 80\% of similarity with experimental real data, we perform these steps again until the similarity is 80\% or higher. We select the 80\% similarity threshold as a good trade-off between the run-time (to generate the fault models) and accuracy (of the generated fault models). As shown in Figure~\ref{fig:simi_threshold}, if we increase the threshold level, the run time of MoRS increases drastically. However, increasing the threshold is not achieve significant accuracy compared to the optimal threshold. Below this threshold, the approximate model converges to the Random Model that do not have an acceptable accuracy.}

%% file: sections/04-ex_methodology.tex
\section{Experimental Methodology}

MoRS is a general framework that generates approximate fault maps for undervolted SRAM blocks. In this study, we test MoRS on state-of-the-art Deep Neural Networks.
Our experiments are based on injecting faults into weights of trained DNNs. To evaluate how precise MoRS is we use Caffe\cite{jia2014caffe}. Also, we perform different quantizations (precisions), bit-mappings, and value masking to diversify our experiments. The summary of these options is in Table \ref{tab:options}. 

\begin{table}[!h]
\caption{Different options for evaluation}
\label{tab:options}
\resizebox{\linewidth}{!}{%
\begin{tabular}{|c|l|}
\hline
\textbf{Options}                        & \textbf{Name}                           \\ \hline
\multirow{4}{*}{\textbf{Precision}}     & 32-bit single-precision floating point \\ \cline{2-2} 
                                        & \revA{16-bit half-precision floating point}  \\ \cline{2-2} 
                                        & 8-bit fixed point (Q4.4)               \\ \cline{2-2} 
                                        & 4-bit fixed point (Q2.2)               \\ \cline{2-2} 
                                        & 1-bit (Binary)                          \\ \hline
\multirow{4}{*}{\textbf{Bit-Mapping}}   & MSB                                     \\ \cline{2-2} 
                                        & LSB                                     \\ \cline{2-2} 
                                        & First half MSB and other half LSB       \\ \cline{2-2} 
                                        & First half LSB and other half MSB       \\ \hline
\multirow{2}{*}{\textbf{Value Masking}} & Infinity or NaN to 1                    \\ \cline{2-2} 
                                        & Infinity or NaN to 0                    \\ \hline
\end{tabular}
}
\end{table}

\begin{figure}[h]
  \centering
  \includegraphics[width=\linewidth]{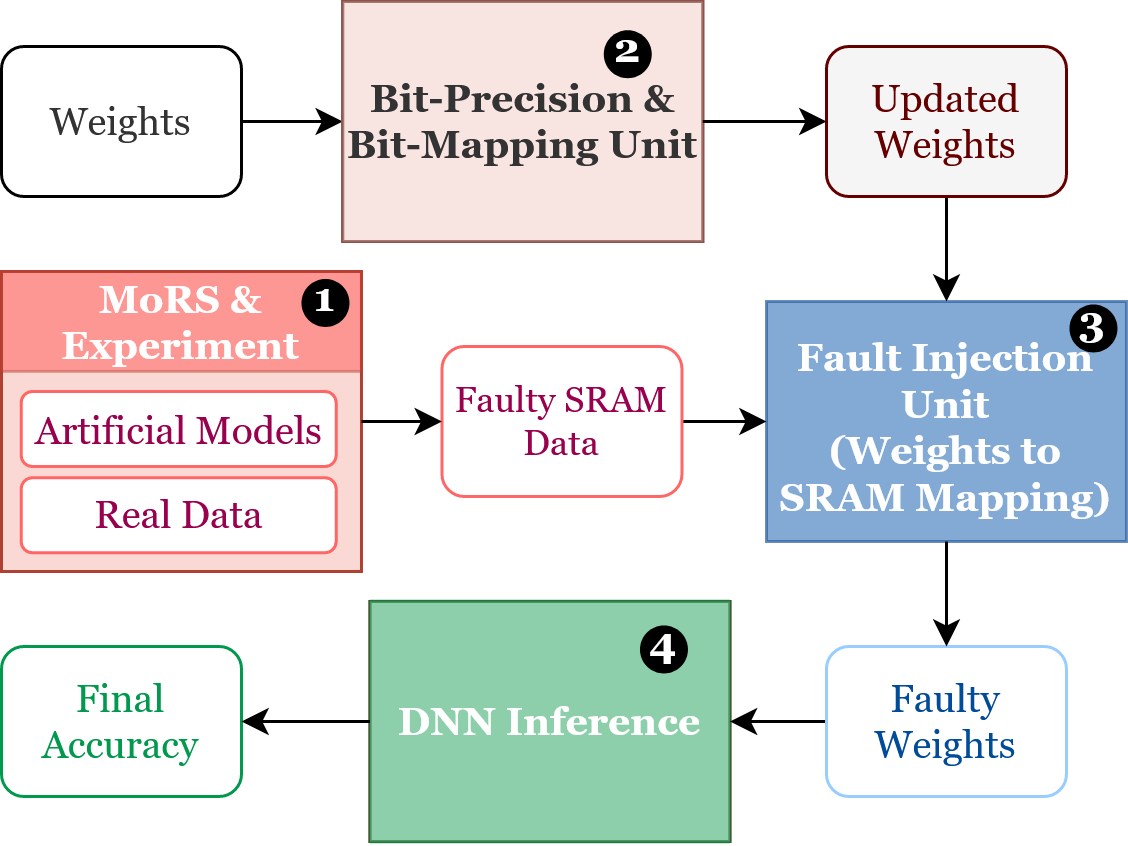}
  \caption{Overall Methodology}
  \label{fig:eval_fig}
\end{figure}

The experiment is performed for each voltage level between $V_{min}$ and $V_{crash}$, precision, mapping, and masking option. To compare artificial models with real data we also process this methodology for real data. Artificial Models are Random Model and Mixed Model. Random Model is a naive random baseline used in prior works to inject faults. Mixed Model is an approximate model, the output of the MoRS. Real Data is the experimental data\cite{salami2018bram} extracted from the prior empirical study\cite{salami2018comprehensive}. To evaluate real data, we select the required amount of SRAM blocks randomly from real data. Instead of Artificial Models, we process real data to evaluate in Figure \ref{fig:eval_fig}.

The methodology consists of four parts as shown in Figure \ref{fig:eval_fig}: \ding[1.5]{182} MoRS \& Experiment, \ding[1.5]{183} Bit-Precision \& Bit-Mapping Unit, 3 \ding[1.5]{184} Fault Injection Unit (Weights to SRAM Mapping) and \ding[1.5]{185} DNN Inference. In the first step, we acquire artificial models from MoRS that are explained in Section 3. In the second step, the Fault Injection Unit changes healthy weights into updated weights by performing mapping and precision options. In the third step, with the outputs of the first and second steps, we generate faulty weights. In the last step, we obtain the final accuracy percentage.

\textbf{1. MoRS \& Experiment.} In this step, we choose which SRAM data is sent to Fault Injection Unit. 
To evaluate MoRS, we generate a baseline model, \textit{i.e.,} Random Model. 
Random Model is generated by uniform random distribution with coarse-grained features. 
We process every step in MoRS described in Section II. 
However, we apply a directly random distribution function to only coarse-grained features instead of applying custom probabilistic modeling and algorithm to coarse-grained and fine-grained features.

In Random Model, first, with a given number of SRAM blocks we determine which and how many blocks are faulty by using $P_{S}$ value. 
Then, using $P_{F}$ value we calculate how many cells are going to be faulty. 
When it is calculated, we randomly inject faults in cells of randomly selected faulty SRAM blocks. 
Because of the uniform random distribution, every cell in faulty blocks has the same probability. 
The differences between Mixed Model and Random Model are summarized in Table \ref{tab:feat}.

In Figure \ref{fig:eval_fig}, we call Mixed Model and Random Model Artificial Models. To understand how accurate our approximate model we evaluate empirical data also called Real Data. Since evaluated networks utilize 850 SRAM blocks at maximum, we randomly choose the required amount of SRAM blocks.

\textbf{2. Bit-Precision \& Bit-Mapping Unit.} Quantization and undervolting are both effective techniques to improve the energy efficiency of DNNs. However, they may lead to accuracy loss with aggressive exploitation. MoRS enables us to explore their correlation to find an optimal operating point.

In this step, we change weights according to precision and mapping options. Caffe's weights are 32-bit single-precision floating points. Since each row of SRAM block has 16-bit, we store those weights in two rows when precision is not reduced. 

\revA{We use four fixed point precisions: 16-bit half-precision floating point, 8-bit (Q.4.4), 4-bit (Q2.2), and binary}. To enable fixed-precision options we use a prior study\cite{milde2017adaptation} an adapted version of the original Caffe with limited numerical precision of weights.\revA{ For 16-bit half-precision floating point, we use NVCaffe~\cite{nvidiacaffe}, NVIDIA-maintained Caffe that supports 16-bit half-precision floating point train and inference.} When precision is reduced to $X-bit$, we store $16/X$ weights in one row consecutively. Therefore, by reducing precision, the usage of SRAM blocks and power consumption decrease with a cost of accuracy loss.

In addition to precision options, we change the mapping of weights to SRAM blocks. There are four mapping options: MSB, LSB, the first half of bit MSB, and another half of bits LSB, the first half of bits LSB, and another half of bit MSB. MSB means the most significant bit of weights maps to the first cell of a row whereas LSB means the least significant bit of weights maps to the first cell of a row.

\textbf{3. Fault Injection Unit (Weights to SRAM Mapping).} After precision and mapping options we obtain updated weights. In this step, we use artificial models to inject faults in updated weights. Each cell of artificial models contains either faulty or healthy information. If a bit of weight is mapped in the faulty cell, we flip its value. Else, the value is not changed. When this bit-flip operation performs, sometimes the value of weights could be infinity or NaN. To prevent this situation we mask these values to either one or zero. \revA{The masking operation is only performed for 32-bit single-precision and 16-bit half-precision floating point. Because fixed point does not have any mantissa or exponent parts to converge NaN or infinite value. In our study, the largest fixed point representation is $Q_{4.4}$ and its maximum value is 15.  }

\textbf{4. DNN Inference.} After injecting faults to the weights, we use Caffe Framework to measure the accuracy of neural networks. Our most accurate baseline is Real Data. We train our model based on part of real data and we test it using another of that data. To diversify we have four different bit-mapping, three different precision, and two different masking options for each voltage level.

%% file: sections/05-ex_results.tex
\begin{figure*}[ht]
\captionsetup[subfigure]{justification=centering}
\centering
\subfloat[$Infinity-NaN \ value \leftarrow 0$ $Mapping \leftarrow MSB$][$Infinity-NaN \ value \leftarrow 0$ \\ $Mapping \leftarrow MSB$]{{\includegraphics[width=0.25\linewidth]{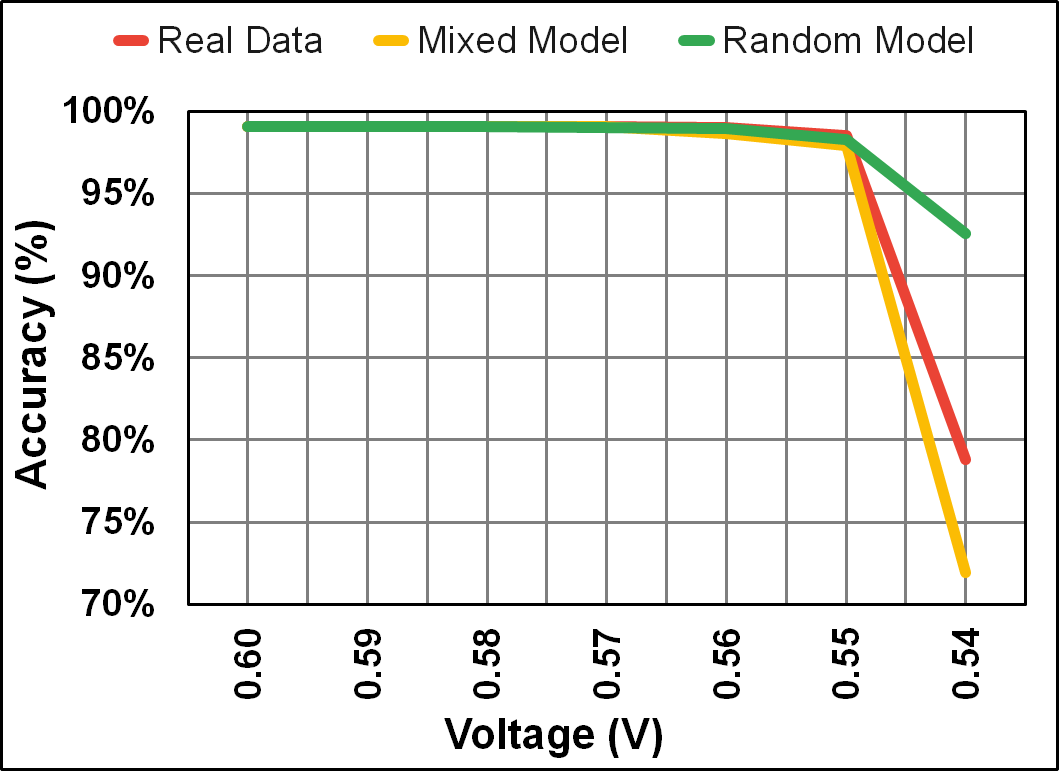} }\label{subfig:i0_mmsb}}%
\subfloat[$Infinity-NaN \ value \leftarrow 1$ $Mapping \leftarrow MSB$][$Infinity-NaN \ value \leftarrow 1$ \\ $Mapping \leftarrow MSB$]{{\includegraphics[width=0.25\linewidth]{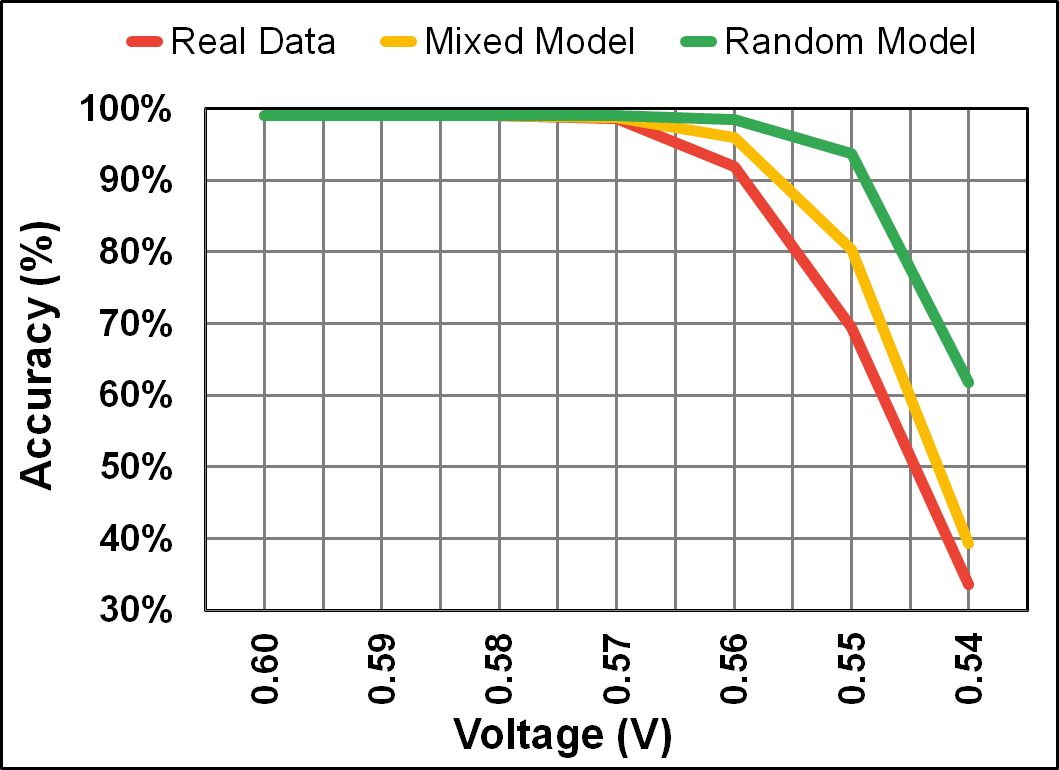} }\label{subfig:i1_mmsb}}%
\subfloat[$Infinity-NaN \ value \leftarrow 0$ $Mapping \leftarrow LSB$][$Infinity-NaN \ value \leftarrow 0$ \\ $Mapping \leftarrow LSB$]{{\includegraphics[width=0.25\linewidth]{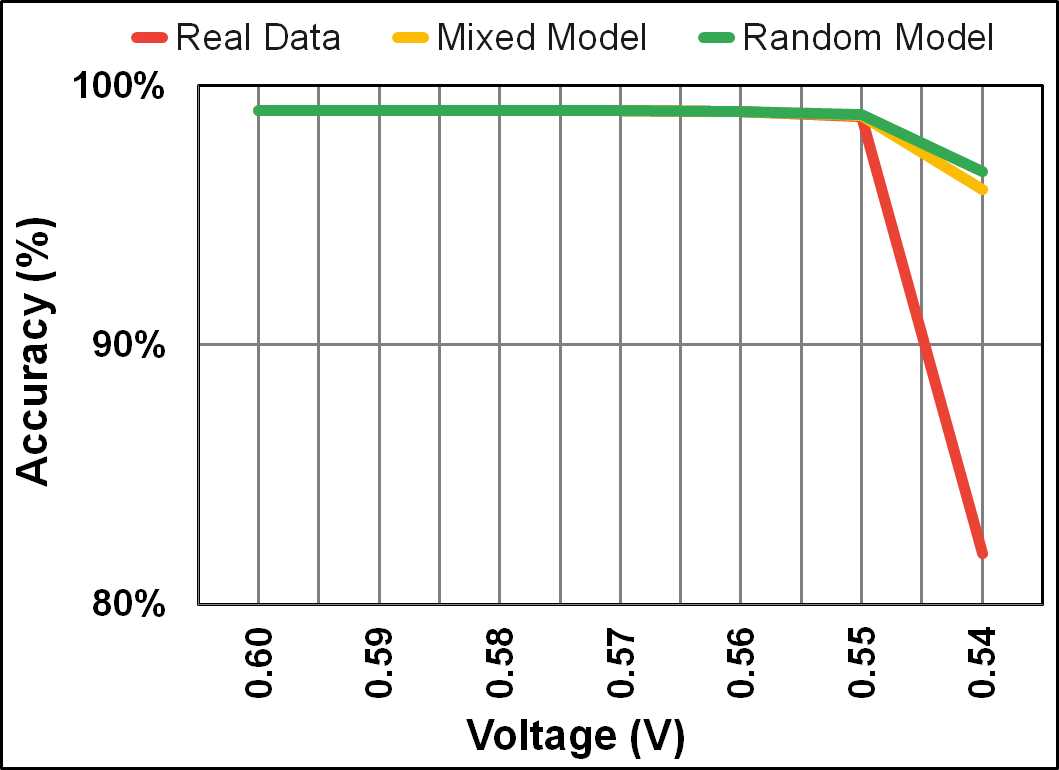} }\label{subfig:i0_mlsb}}%
\subfloat[$Infinity-NaN \ value \leftarrow 1$ $Mapping \leftarrow LSB$][$Infinity-NaN \ value \leftarrow 1$ \\ $Mapping \leftarrow LSB$]{{\includegraphics[width=0.25\linewidth]{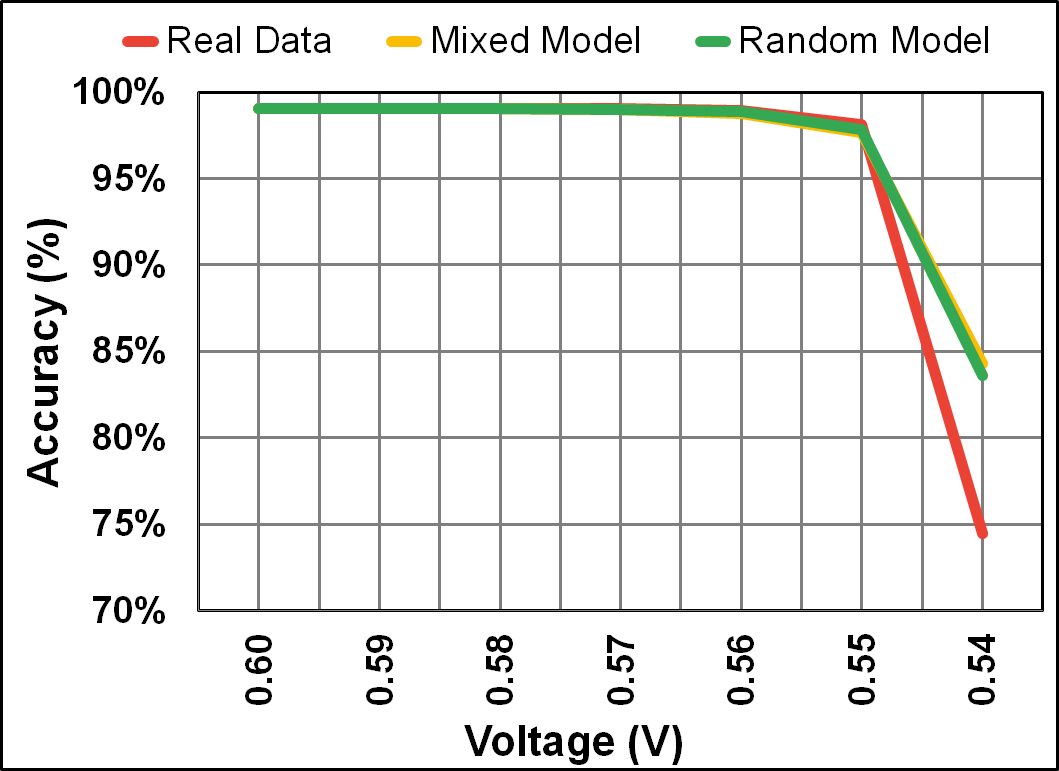} }\label{subfig:i1_mlsb}}%
\\
\subfloat[$Infinity-NaN \ value \leftarrow 0$ $Mapping \leftarrow MSB\mid LSB$][$Infinity-NaN \ value \leftarrow 0$ \\ $Mapping \leftarrow MSB\mid LSB$]{{\includegraphics[width=0.25\linewidth]{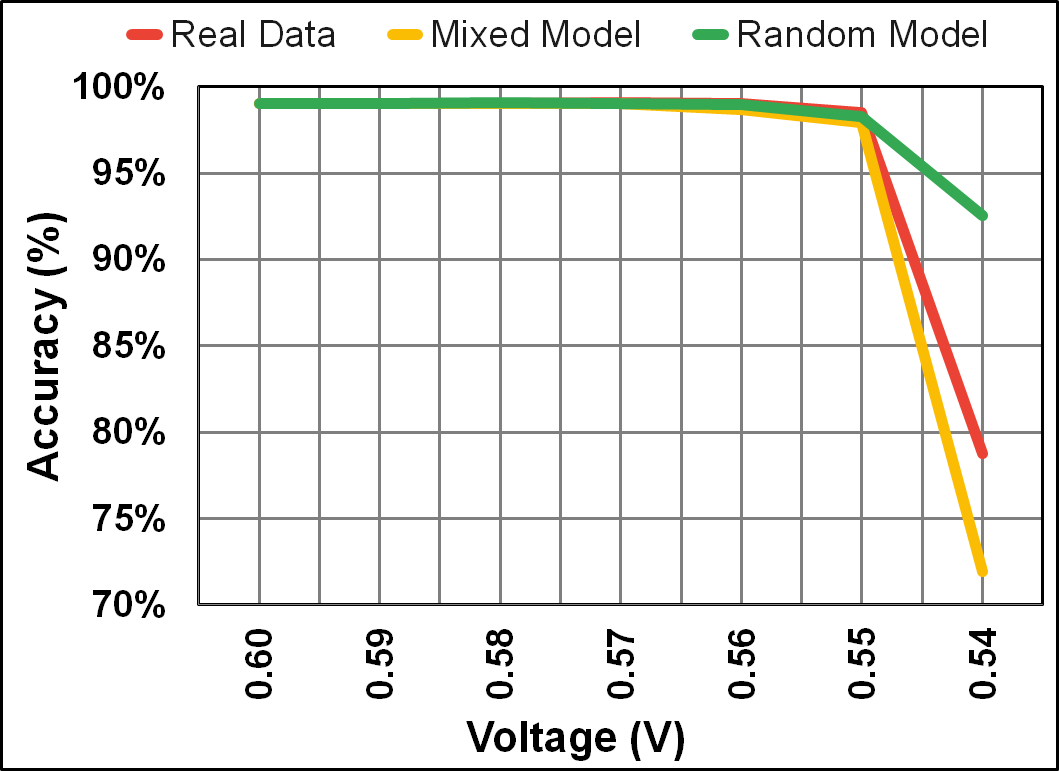} }\label{subfig:i0_mmsblsb}}%
\subfloat[$Infinity-NaN \ value \leftarrow 1$ $Mapping \leftarrow MSB\mid LSB$][$Infinity-NaN \ value \leftarrow 1$ \\ $Mapping \leftarrow MSB\mid LSB$]{{\includegraphics[width=0.25\linewidth]{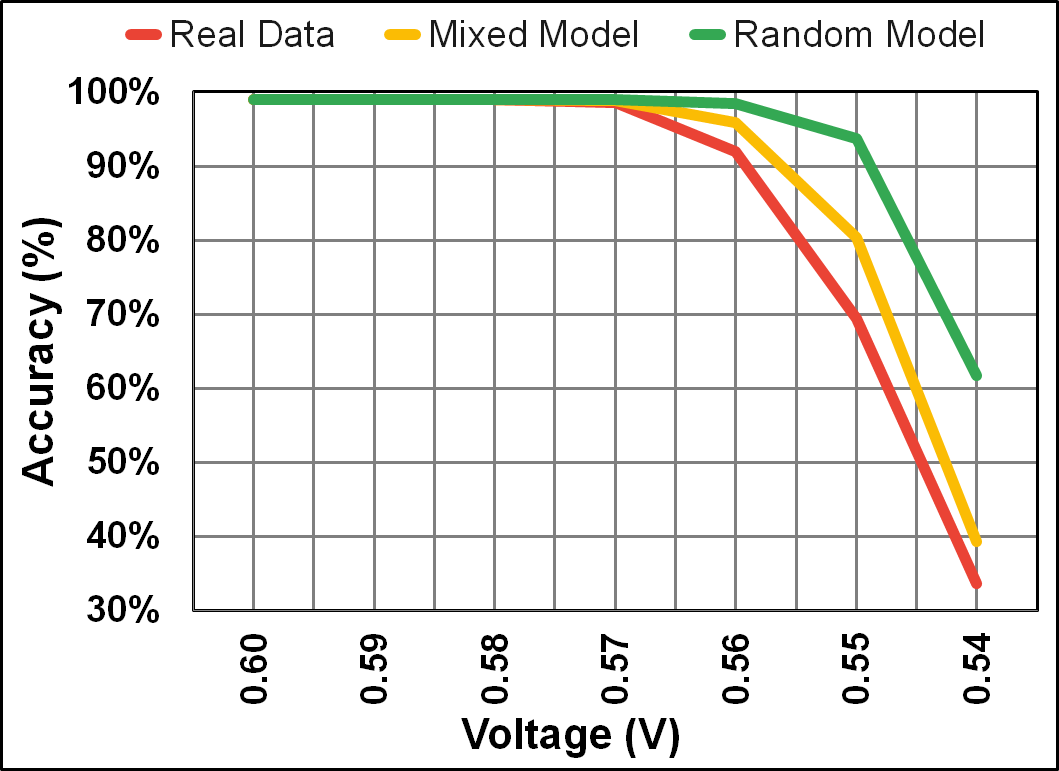} }\label{subfig:i1_mmsblsb}}%
\subfloat[$Infinity-NaN \ value \leftarrow 0$ $Mapping \leftarrow LSB\mid MSB$][$Infinity-NaN \ value \leftarrow 0$ \\ $Mapping \leftarrow LSB\mid MSB$]{{\includegraphics[width=0.25\linewidth]{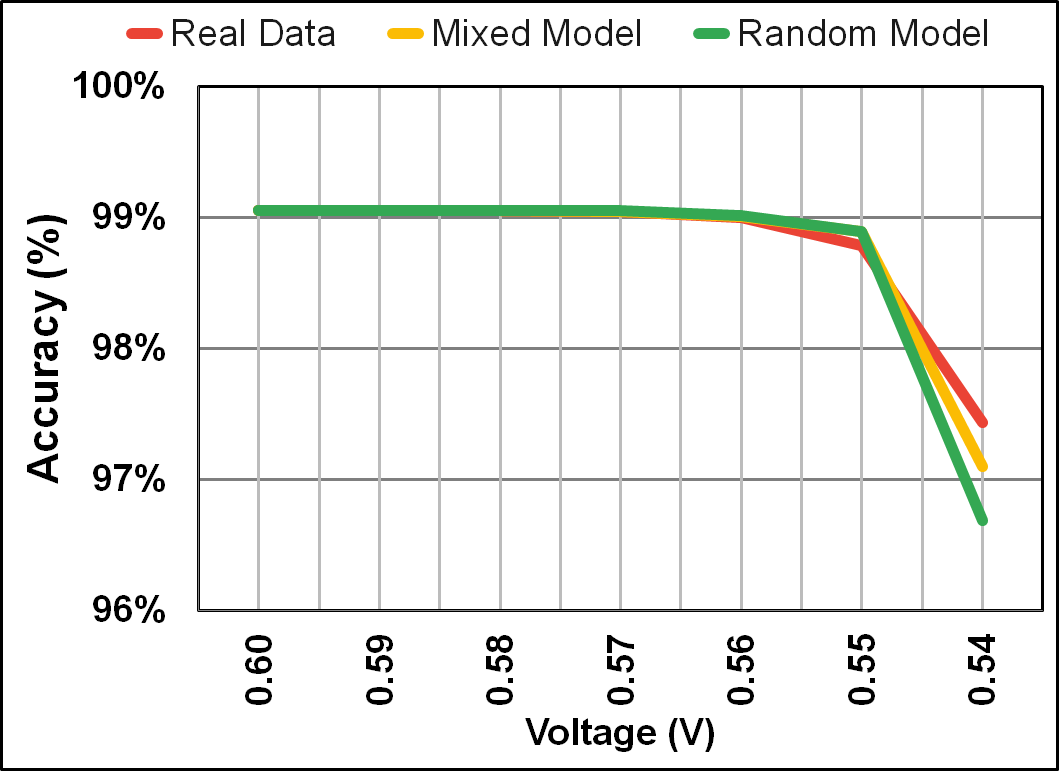} }\label{subfig:i0_mlsbmsb}}%
\subfloat[$Infinity-NaN \ value \leftarrow 1$ $Mapping \leftarrow LSB\mid MSB$][$Infinity-NaN \ value \leftarrow 1$ \\ $Mapping \leftarrow LSB\mid MSB$]{{\includegraphics[width=0.25\linewidth]{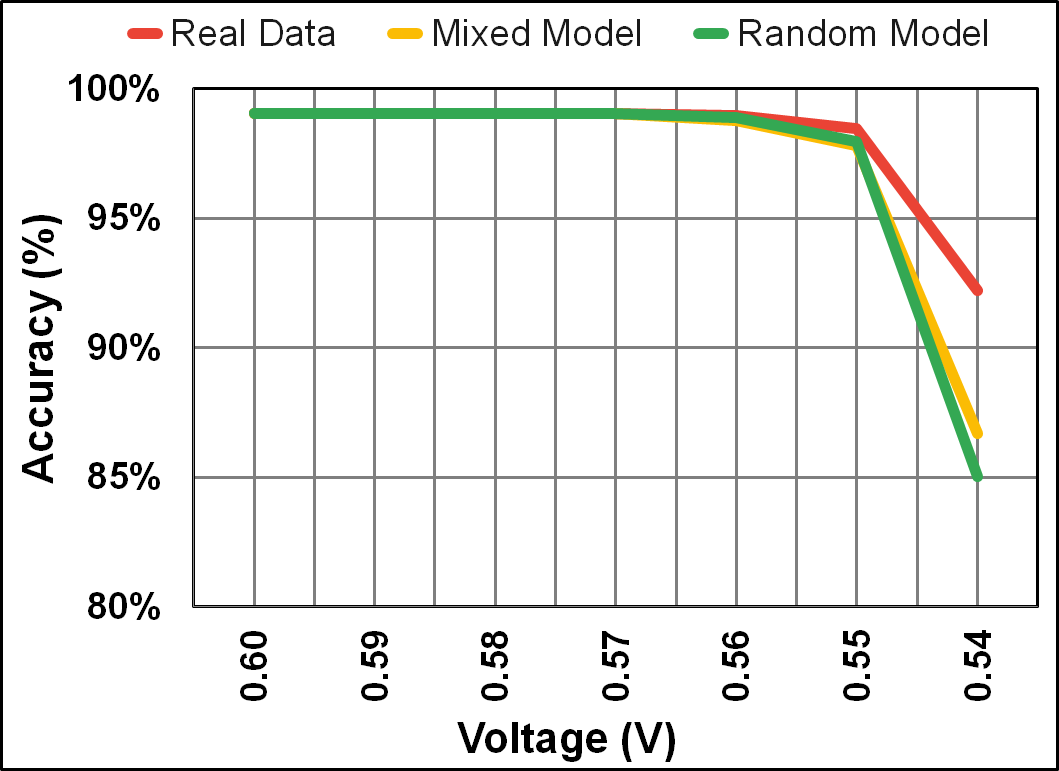} }\label{subfig:i1_mlsbmsb}}%
\caption{Voltage and resilience behavior of artificial models and experimental (real) data on LeNeT-5\cite{lecun1998lenet} network for each bit-mapping and value masking option when precision is not reduced (\revA{32-bit single-precision floating point.})}
\label{fig:options_lenet}
\end{figure*}
\section{Experimental Results}
As we mentioned in the previous section we use Caffe\cite{jia2014caffe}, a deep learning framework, and test the output of MoRS and random fault injection model on two different neural network models: LeNeT-5\cite{lecun1998lenet} with MNIST dataset\cite{mnist} and cuda-convnet\cite{krizhevsky2014cuda} with CIFAR-10 dataset\cite{cifar10}. We perform different bit-mappings and value maskings for each neural network architecture. In addition to these tests, we also perform reduced precision tests on LeNeT-5. Details of each evaluated benchmark are summarized in Table \ref{tab:benchmarks}.

\begin{table}[]
\caption{Details of evaluated neural networks}
\label{tab:benchmarks}
\resizebox{\linewidth}{!}{%
\begin{tabular}{|l|c|c|c|}
\hline
\textbf{NN Model}                   & LeNeT-5\cite{lecun1998lenet} & cuda-convnet\cite{krizhevsky2014cuda} \\ \hline
\textbf{Dataset Name}               & MNIST\cite{mnist}   & CIFAR-10\cite{cifar10}     \\ \hline
\textbf{\# of Weights}              & 430500  & 89440        \\ \hline
\textbf{\# of SRAM Blocks Utilized} & 850     & 180          \\ \hline
\textbf{Inference Accuracy (\%)}    & 99.05\% & 79.59\%      \\ \hline
\end{tabular}
}
\end{table}

\subsection{Overall Resilience}

Figure \ref{fig:options_lenet} shows the accuracy of LeNeT-5\cite{lecun1998lenet} on the MNIST dataset\cite{mnist} with different bit-mapping and value masking options. We observe that in all options Mixed Model is more precise than Random Model. Also, if the application becomes less resilient,  Mixed Model is closer to the real data and the gap between the baseline and Mixed Model is increasing in terms of how close they are to the real data. 

We observe that that if we mask infinity and NaN value to 0, the LeNeT-5 network is more resilient than masking to 1. We think that the cause of this situation is in the MNIST dataset hand-written digits are represented by one and the rest of the background is represented by zero. Therefore, ones more impact than zeros when it comes to classification. 

We see that MSB and $MSB\mid LSB$ mapping cause more faults and are less resilient than LSB and $LSB\mid MSB$ mapping. We find that the cause of this is undervolting-based faults generally occur in first cells. Since MSB means the most significant bit, when bit-flips happen it affects the corresponding value more than others. Figure \ref{subfig:i1_mmsb} and Figure \ref{subfig:i1_mmsblsb} show more characteristic behavior and are less resilient than others since both have MSB mappings and masking to 1 option. 

In Figure,\ref{fig:options_lenet} we perform 300 iterations for each option (value masking, bit mapping). Then we average these options for LeNeT-5 and cuda-convnet\cite{krizhevsky2014cuda} network architectures. In addition to these, we perform these experiments at different precision levels for LeNeT-5 and for each precision level, we average all options. 

\begin{figure}[ht]
  \centering
  \includegraphics[width=0.8\linewidth]{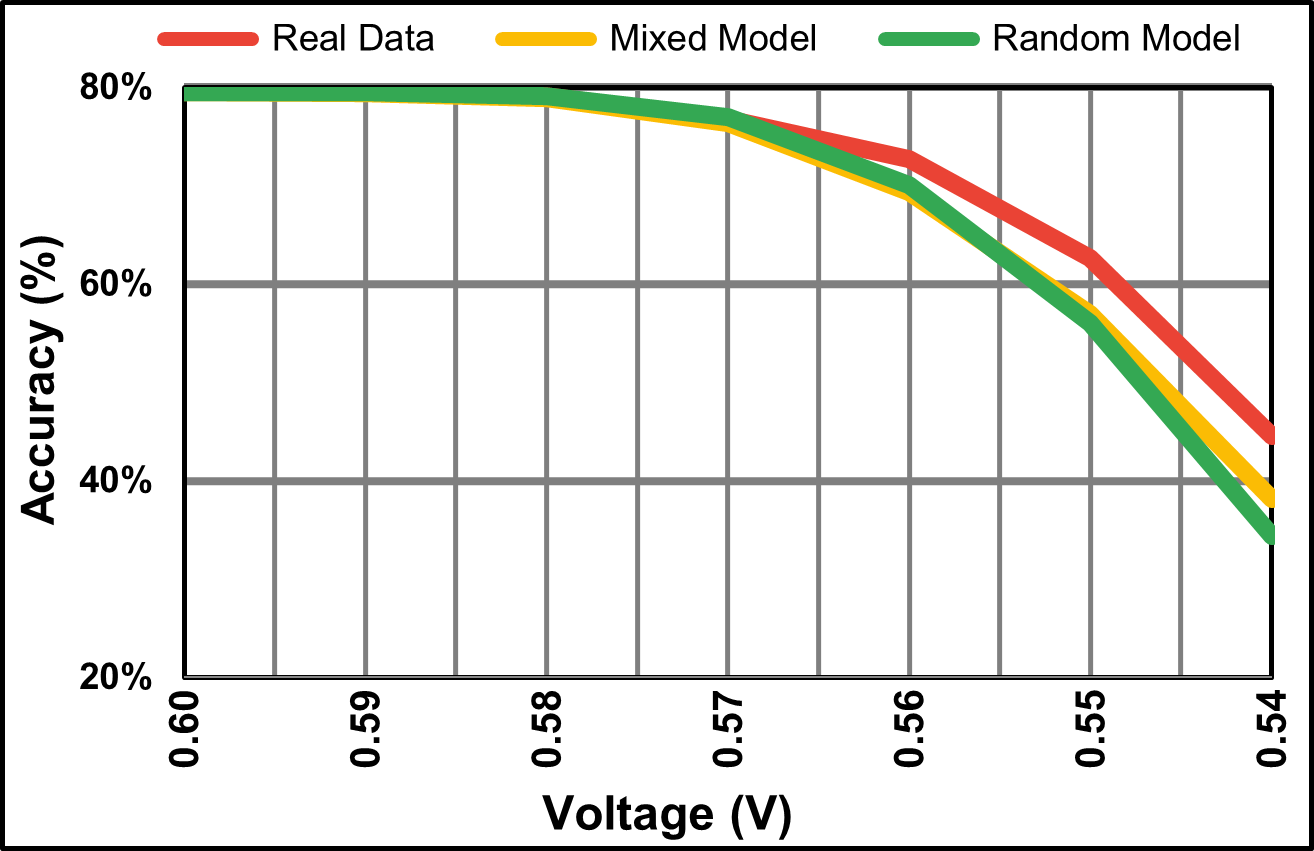}
  \caption{Average of all options on voltage and resilience behaviour of artificial models and experimental (real) data on cuda-convnet network without reducing precision.}
  \label{fig:cifar10}
\end{figure}

Figure \ref{fig:cifar10} shows the accuracy of cuda-convnet on the CIFAR-10 dataset\cite{cifar10} without reducing precision, \textit{i.e,} the precision of weights is 32-bit floating points. 
\revA{As we evaluate the LeNeT-5\cite{lecun1998lenet} network in Figure \ref{subfig:lenet_32bits}, we average every bit-mapping and value masking option for cuda-convnet \cite{krizhevsky2014cuda} network.}

We find that the cause of LeNeT-5 works better is cuda-convnet on CIFAR-10 is rather smaller in terms of the number of weights and SRAM block utilization. 
However, even in this network, Mixed Model is 1.47x more accurate than Random Model in terms of accuracy difference with Real Data on average.
While the supply voltage of SRAM blocks is reducing, we observe more characteristic and distinguishing behavior from real data. 
At $V_{crash}$ level, the accuracy difference between Mixed Model and Real Data is only 6\% whereas between Random Model and Real Data is 10\%.

\subsection{Quantization}

\begin{figure*}[!h]
\centering
\captionsetup[subfigure]{justification=centering}
\subfloat[32-bit single-precision floating point][32-bit single-precision \\ floating point]{{\includegraphics[width=0.2\linewidth]{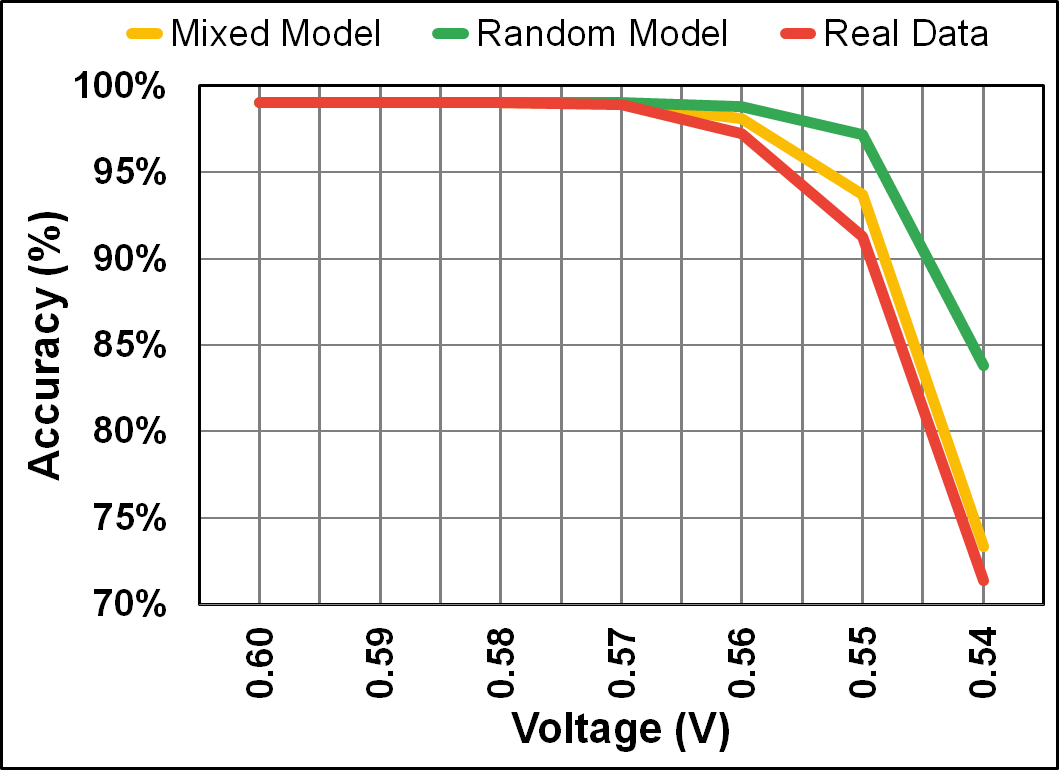} }\label{subfig:lenet_32bits}}%
\subfloat[16-bit half-precision floating point]{{\includegraphics[width=0.2\linewidth]{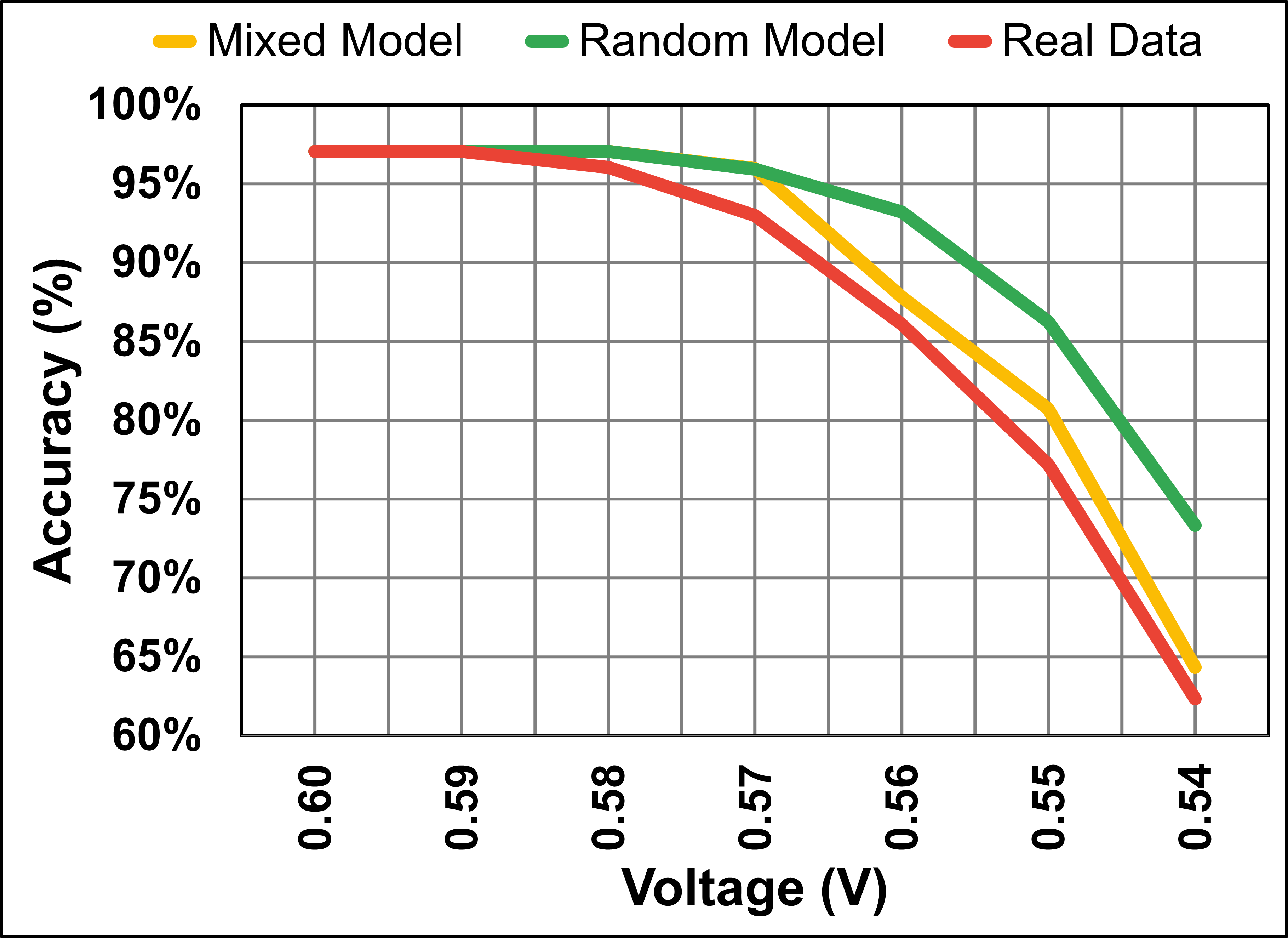} }\label{subfig:lenet_16bits}}
\subfloat[8-bit fixed point (Q4.4)]{{\includegraphics[width=0.2\linewidth]{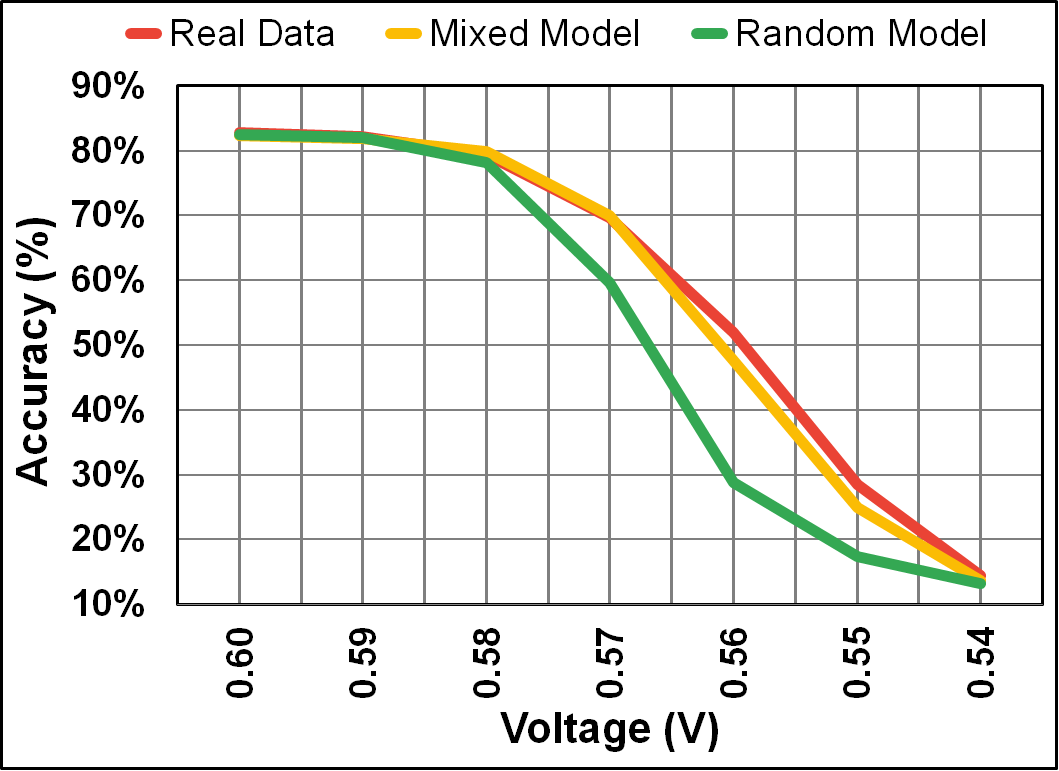} }\label{subfig:lenet_8bits}}
\subfloat[4-bit fixed point (Q2.2)]{{\includegraphics[width=0.2\linewidth]{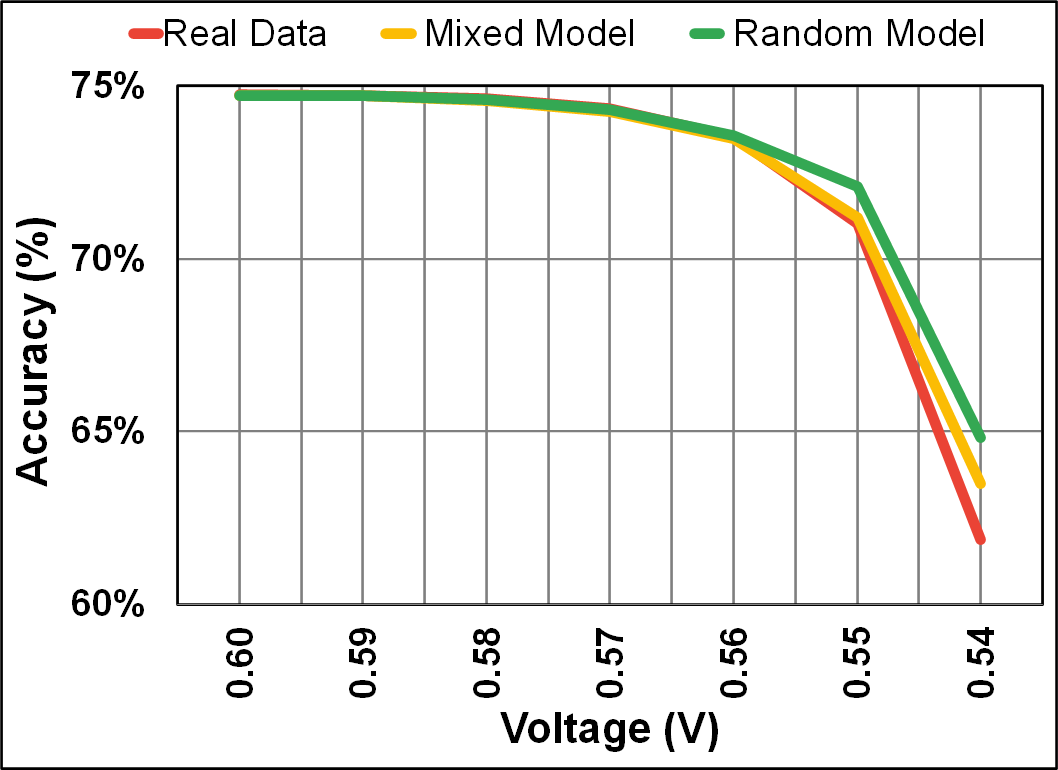} }\label{subfig:lenet_4bits}}
\subfloat[1-bit (Binary)]{{\includegraphics[width=0.2\linewidth]{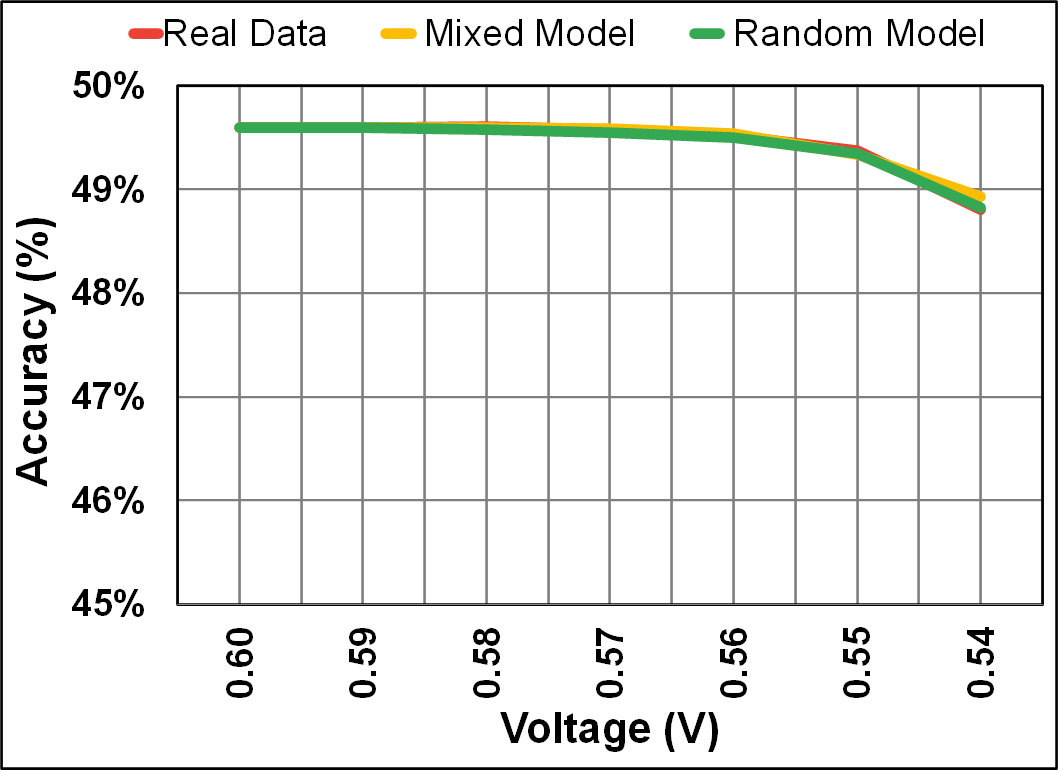} }\label{subfig:lenet_1bit}}
\caption{ \revA{Average of all options on voltage and resilience behaviour of artificial models and experimental (real) data on LeNeT-5 network with different precisions.} }
%\Description{Average LeNeT-5 behaviour for all data.}
\label{fig:lenet_avg}
\end{figure*}

Figure \ref{fig:lenet_avg} shows the accuracy of LeNeT-5 on the MNIST dataset at different precision levels. We see that in each precision level our the Mixed Model has more similar behavior than Random Model has to the real data. Figure \ref{subfig:lenet_32bits} is the average of all options demonstrated in Figure \ref{fig:options_lenet}. 

Figure~\ref{subfig:lenet_32bits} shows the resilience of both artificial models and real data under nominal precision level (32-bit single-precision floating point) in terms of accuracy. Mixed Model is 3.74x closer than Random Model to the real data on average. At $V_{crash}$, the accuracy difference between Mixed Model and Real Data is 2.46\%, on the other hand, the difference between Random Model and Real Data is 12.47\%. 

\revA{Figure \ref{subfig:lenet_16bits} shows the accuracy of LeNeT for weights with the precision reduced to 16-bit half-precision floating point. The behavior of all models are similar to the 32-bit architecture, since 16-bit precision is mostly cover all values in 32-bit~\cite{micikevicius2017mixed,markidis2018nvidia} without significant affect on accuracy. The highest accuracy difference between real data and Mixed Model is 4.47\% at 0.55V whereas the highest accuracy difference between real data and Random Model is 9.05\% at 0.57V. The average accuracy difference of all voltage level between real data and Mixed Model is 2.25\% whereas between real data and Random Model is 4.84\%, which is 2.15x worse than Mixed Model in terms of similarity of the behavior of real data.}

Figure~\ref{subfig:lenet_8bits} shows the accuracy of LeNeT with 8-bit precision for weights. We see that compared to 32-bit, the network are less resilient to undervolting faults. On average, Mixed Model is 7x better than Random Model in terms of approximation to real data. At $560mV$, the accuracy gap between Mixed Model and Real Data is 4\% while the difference between Random Model and Real Data is 23\%. 

Figure~\ref{subfig:lenet_4bits} shows the accuracy of both artificial models and real data when the precision of weights reduces to 4-bit. We see that reducing the precision of weights 8-bit to 4-bit network become more resilient and more fault-tolerant to errors based on voltage underscaling. Although, at $V_{nom}$, 8-bit LeNeT is more accurate than 4-bit LeNeT, at $V_{crash}$, 8-bit accuracy is 14\% whereas 4-bit accuracy is 62\%. Even in this unexpected situation, artificial models of the MoRS framework have similar behavior to real data. On average, Mixed Model is 2x closer to real data than Random Model. At $550mV$, the difference in accuracy between Mixed Model and real data is around 1.5\%, while the difference in accuracy between Random Model and real data is 3\%. 

Figure~\ref{subfig:lenet_1bit} shows the accuracy of LeNeT when precision reduces to 1-bit. In 1-bit tests, we map weights to three different value sets. First one is \{-1,1\}, second is \{-1,0\} and the last one is \{0,1\}. However, we have not observed much difference between value sets. As we see in 4-bit, 1-bit LeNeT network becomes more resilient to faults. The difference between $V_{nom}$ accuracy and $V_{crash}$ accuracy is 0.79\%. Both artificial models have the same behavior and do not have significant difference in accuracy. Mixed Model has 0.03\% difference in accuracy, whereas Random Model is 0.02\%. Because of these negligible statistics, we do not add 1-bit LeNeT to Figure \ref{fig:comparison}. 

To point out the resilience of reduced precision networks, we examine the accuracy drop between $V_{nom}$ and $V_{crash}$. The drop is 68.4\%, 12.9\% and 0.79\% for 8-bit, 4-bit and 1-bit LeNet, respectively.

\begin{table}[h]
\begin{tabular}{|c||c|c|}
\hline
\textbf{LeNeT-5 Precision} & \multicolumn{1}{l|}{\textbf{SRAM Utilization}} & \multicolumn{1}{l|}{\textbf{ Accuracy}} \\ \hline\hline
\textbf{32-bit FP}      & 850 & 99.05\% \\ \hline
\textbf{16-bit Half FP} & 425 & 97.03\%\\ \hline
\textbf{8-bit (Q4.4)}   & 213 & 82.79\% \\ \hline
\textbf{4-bit (Q2.2)}   & 107 & 74.75\% \\ \hline
\textbf{1-bit (Binary)}  & 27  & 49.59\% \\ \hline
\end{tabular}
\caption{ \revA{LeNeT-5 SRAM Block Utilization and Inference Accuracy(at $V_{nom}$) under different precisions}}
\label{tab:precision_table}
\end{table}

\revA{Table \ref{tab:precision_table} shows the SRAM utilization and accuracy in various precision levels of weights. Inference accuracy represents the accuracy at nominal voltage level ($V_{nom}$). 32-bit FP denotes the 32-bit single-precision floating point, 16-bit Half FP stands for 16-bit half-precision floating points.}

\subsection{Comparison of Artificial Models}

\begin{figure}[!h]
  \centering
  \includegraphics[width=\linewidth]{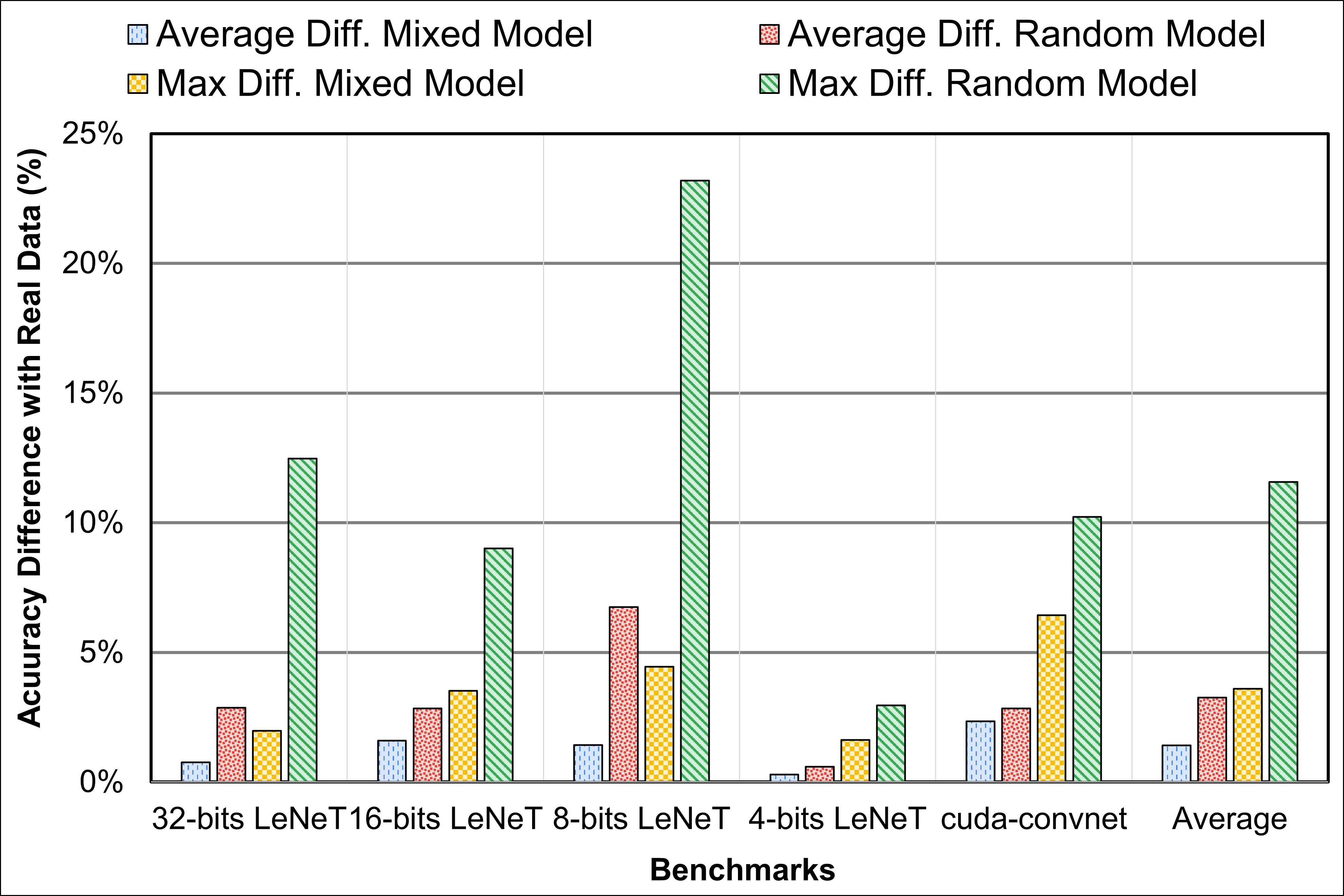}
  \caption{ \revA{The accuracy difference between Real Data and each artificial model for all benchmarks.}}
  \label{fig:comparison}
\end{figure}

Figure~\ref{fig:comparison} shows the accuracy gap between Real Data and both two artificial models. We see that on the average of all benchmarks, Mixed Model is 3.21x closer than Random Model on average. For most of the benchmarks, the maximum difference in accuracy between Mixed Model and Real Model is under 5\%. However, for the Random Model, the maximum difference in accuracy is 23.2\%.

We conclude that the proposed model not only has the same behavior with real data against undervolting effects but also can imitate real data with a tolerable difference in terms of accuracy. Most importantly, if the system is not resilient to faults, randomized fault injection does not show the behavior of real data. To be precise in how undervolting affects systems, it has to profile the real data in fine-grained. Coarse-grained features are insufficient to model real SRAM behavior when undervolting is performed.

%% file: sections/06-related_work.tex
\section{Related Work}
To the best of our knowledge, this study provides the first approximate fault modeling framework and injection in voltage underscaled SRAMs. In this section, we discuss related work on fault injection and modeling on undervolted systems and the resilience of DNNs. 

\textbf{Resilience of DNNs.} DNNs are inherently reliable to faults. However, in harsh environments, process variations,  voltage underscaling can cause significant accuracy loss. Reagen et al.\cite{reagen2016minerva} propose Minerva a fault mitigation mechanism to mitigate low-voltage SRAM faults effects in DNN accelerators. Salami et al.\cite{salami2018comprehensive} study undervolting SRAM-based FPGA on-chip memories and present intelligently-constrained BRAM placement to mitigate undervolting faults in the NN classification layer. Torres-Huitzil et al.\cite{torreshuitzil2017fault} present a comprehensive review on fault and error tolerance in neural networks and mention mitigation and fault injection techniques. They also mention that more realist/novel fault models need to be developed to understand the effects of faults on neural networks deeply as an open challenge. Deng et al.\cite{deng2015retraining} present a retraining-based mitigation technique for neural networks to become more resilient.  

\textbf{Fault injection.} Fault injection is a widely used technique in resilience studies. Also, fault injection is used as bit-flip-based adversarial weight attacks also called bit-flip attacks(BFA)\cite{rakin2021tbfa,he2020defending,rakin2019bitflip} and fault-injection attacks\cite{liu2017fault,kelly2017characterising}. Many studies focus on the reliability and resilience of systems on soft-errors, noise\cite{li2017understanding,neggaz2018areliability,reagen2018ares,salavati2012multi}, and voltage underscaling by injecting faults. Koppula et al.\cite{koppula2019eden} propose a framework, EDEN, that proposes combining training on profiled DRAM faults in terms of timing violations and voltage underscaling with mitigation strategies and mappings. EDEN provides four different error models since uniform random distribution does not cover whole DRAMs. Chatzidimitrou et al.\cite{chatzidimitriou2019assessing,chatzidimitrou2018analysis} inject faults randomly to branch prediction units to examine the effects of voltage underscaling.  Chandramoorthy et al.\cite{chandramoorthy2019resilient} study the undervolting faults in different layers of networks by injecting faults to SRAMs randomly and do not take account of the patterns or spatial distribution of bit errors. Stutz et al.\cite{stutz2020bit} propose random bit error training assumed voltage underscaled SRAMs faults distribute randomly. Salami et al. \cite{salami2018ontheresilience} study the resilience of RTL NN Accelerators and fault characterization and mitigation. To characterize and mitigate they assume that each bitcell of SRAMs has the same probability. Yang et al. \cite{yang2017sram} study energy-efficient CNNs by performing voltage scaling on SRAMs. To study the effect of bit errors they hypothesize that errors in SRAM are roughly uniformly distributed. The prior work\cite{ganapathy2017oncharacterizing} on Near-Threshold Voltage FinFET SRAMs presents a fault model for SRAMs based on uniform random distribution. Givaki et al.\cite{givaki2020resilience} study the resilience of DNNs under reduced voltage SRAM-based FPGA on-chip memories by using directly the experimental data to examine the training phase of DNNs. 

All of these undervolted on-chip fault injection studies perform injection randomly and do not take into account fine-grained profiling such as spatial distances between cells, row-based and column-based approaches. Randomly injecting faults approach can cause misleading to understand how the system works under the low-voltage domain. As we mentioned in Section 4, if the system does not have much resilience to voltage underscaling, 
the Mixed Model of MoRS is 7x closer than the randomly injected model to the real data.

%% file: sections/07-conclusion.tex
\section{Conclusion}
%SRAMs consume more power as workloads are more dependent on-chip memory like DNNs. To save energy, undervolting is a technique that reduces the supply voltage below the nominal voltage level. However, further underscaling the voltage below the guardband causes timing-related bit-flips. Fault injection techniques are widely used to understand how low-voltage-based faults affect systems. Low-voltage fault injection techniques are performed by randomly injecting faults into random locations.
In this paper, we propose MoRS, a framework that generates the first approximate fault injection model \textit{i.e.,} artificial fault maps. The advantage of the proposed framework is to inject errors into various systems including heterogeneous computing devices. We evaluated the accuracy of the proposed framework for state-of-the-art DNN applications. To evaluate our proposed model, we measure the difference in accuracy between artificial error models and real data. We show that compared to random model-based error injections, the proposed model can provide 3.21x on average closer than to the real data.

%% file: sections/08-ack.tex
\section*{Acknowledgement}
This work is partially funded by Open Transprecision Computing (OPRECOM) project, Summer of Code 2020 and funded by Ministerio de Ciencia e Innovacion - Agencia Estatal de Investigacion (PID2019-107255GB-C21/AEI/10.13039/501100011033).

%% file: sections/08-authors_biography.tex
\begin{IEEEbiography}[{\includegraphics[width=1in,height=1.25in,clip,keepaspectratio]{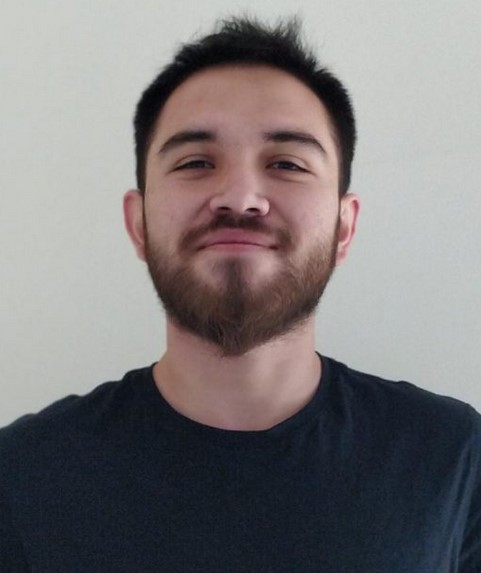}}]{\.{I}smail Emir Y\"{u}ksel}
is an MSc student and researcher in the Computer Engineering Department of TOBB University of Economics and Technology (TOBB ETÜ). He received his BSc in Electrical Electronics Engineering from TOBB University of Economics and Technology in 2019. His research interests are energy-efficient heterogeneous computing and low-power \& fault-resilient hardware accelerators.
\end{IEEEbiography}

\begin{IEEEbiography}[{\includegraphics[width=1in,height=1.25in,clip,keepaspectratio]{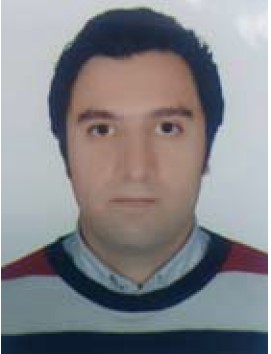}}]{Behzad Salami}
is a post-doctoral researcher in the Computer Science (CS) department of Barcelona Supercomputing Center (BSC) and an affiliated research member of SAFARI Research Group at ETH Zurich. He received his Ph.D. with honors in Computer Architecture from Universitat Polit\'{e}cnica de Catalunya (UPC) in 2018. Also, he obtained MSc and BSc degrees in Computer Engineering from Amirkabir University of Technology (AUT) and Iran University of Science and Technology (IUST), respectively. He has received mutiple awards and grants for his research. His research interests are heterogeneous systems,  low-power \& fault-resilient hardware accelerators, and near-data processing systems. Contact him at: behzadsalami@gmail.com
\end{IEEEbiography}

\begin{IEEEbiography}[{\includegraphics[width=1in,height=1.25in,clip,keepaspectratio]{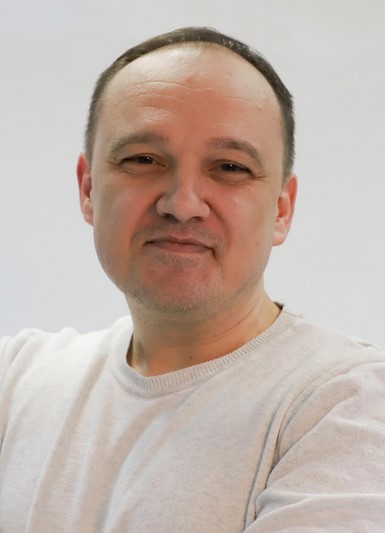}}]{O\u{g}uz Ergin}
 is a professor in the department of computer engineering in TOBB University of Economics and Technology. He received his BS in electrical and electronics engineering from Middle East Technical University, MS, and Ph.D. in computer science from the State University of New York at Binghamton. He was a senior research scientist in Intel Barcelona Research Center prior to joining TOBB ETÜ. He is currently leading a research group in TOBB ETÜ working on energy-efficient, reliable, and high-performance computer architectures.
\end{IEEEbiography}

\begin{IEEEbiography}[{\includegraphics[width=1in,height=1.25in,clip,keepaspectratio]{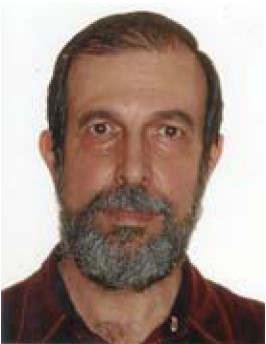}}]{Osman Sabri \"{U}nsal}
is co-manager of the Parallel Paradigms for Computer Architecture research group at Barcelona Supercomputing Center (BSC). He got his B.S., M.S., and Ph.D. in Computer Engineering from Istanbul Technical University, Brown University, and the University of Massachusetts, Amherst respectively. His current research interests are in computer architecture, fault-tolerance, energy-efficiency, and heterogeneous computing. He is currently leading LEGaTO EU H2020 research project on heterogeneous energy-efficiency computing.
\end{IEEEbiography}
\begin{IEEEbiography}[{\includegraphics[width=1in,height=1.25in,clip,keepaspectratio]{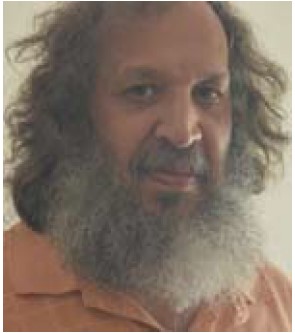}}]{Adri\'{a}n Cristal Kestelman}
received the Licenciatura degree in Computer Science from the Faculty of Exact and Natural Sciences, Universidad de Buenos Aires, Buenos Aires, Argentina, in 1995, and the Ph.D. degree in Computer Science from the Universitat Polit\'{e}cnica de Catalunya (UPC), Barcelona, Spain. Since 2006, he is a co-manager of the Computer Architecture for Parallel Paradigms Research Group at Barcelona Supercomputing Center (BSC). His current research interests include the areas of microarchitecture, multicore, and heterogeneous architectures, and programming models for multicore architectures. Currently, he is leading the architecture development of the vector processor unit in the European Processor Initiative.
\end{IEEEbiography}